\documentclass[final,5p,times,twocolumn]{elsarticle}

\usepackage{graphicx}
\setcitestyle{square}
\usepackage{amssymb}
\usepackage{multicol}
\usepackage{bm}
\usepackage{subfigure}
\usepackage{multirow}
\usepackage{amsmath}
\usepackage{float}
\usepackage{diagbox}
\usepackage{natbib}
\usepackage{hyperref}
\usepackage{color}
\usepackage{lineno}
\usepackage{color}

\journal{NeuroComputing}

\begin{document}

\begin{frontmatter}


\title{Automatic Segmentation of Organs-at-Risk from Head-and-Neck CT using Separable Convolutional Neural Network with Hard-Region-Weighted Loss }



\author[a]{Wenhui Lei}
\author[a]{Haochen Mei}
\author[a]{Zhengwentai Sun}
\author[a]{Shan Ye}
\author[a]{Ran Gu}
\author[a]{Huan Wang}
\author[b]{Rui Huang}
\author[c]{Shichuan~Zhang}
\author[a]{Shaoting~Zhang}
\author[a]{Guotai~Wang\corref{cor}}
\address[a]{School\ of\ Mechanical\ and\ Electrical Engineering, University of\ Electronic\ Science\ and\ Technology\ of\ China, Chengdu, China}
\address[b]{SenseTime Research, Shenzhen, China}
\address[c]{Department\ of\ Radiation\ Oncology, Sichuan Cancer Hospital and Institute, University of Electronic Science and Technology of China, Chengdu, China}
\cortext[cor]{Corresponding author: Guotai Wang.}
\ead{guotai.wang@uestc.edu.cn}

\begin{abstract}
Accurate segmentation of Organs-at-Risk (OAR) from Head and Neck (HAN) Computed Tomography (CT) images with uncertainty information is critical for effective planning of radiation therapy for Nasopharyngeal Carcinoma (NPC) treatment. Despite the state-of-the-art performance
achieved by Convolutional Neural Networks (CNNs) for the segmentation task, existing methods do not provide uncertainty estimation of the segmentation results for treatment planning, and their accuracy is still limited by  the low contrast of soft tissues in CT, highly imbalanced sizes of OARs and large inter-slice spacing. To address these problems, we propose a novel framework for accurate OAR segmentation with reliable uncertainty estimation. First, we propose a Segmental Linear Function (SLF) to transform the intensity of CT images to make multiple organs more distinguishable than existing simple window width/level-based methods. Second, we introduce a novel 2.5D network (named as 3D-SepNet) specially designed for dealing with clinic CT scans with anisotropic spacing. Thirdly, we propose a novel hardness-aware loss function that pays attention to hard voxels for accurate segmentation. We also use an ensemble of models trained with different loss functions and intensity transforms to obtain robust results, which also leads to segmentation uncertainty without extra efforts. Our method won the third place of the HAN OAR segmentation task in StructSeg 2019 challenge and it achieved weighted average Dice of 80.52\% and $95\%$ Hausdorff Distance of 3.043 mm. Experimental results show that 1) our SLF for intensity transform helps to improve the accuracy of OAR segmentation from CT images; 2) With only 1/3 parameters of 3D UNet, our 3D-SepNet obtains better segmentation results for most OARs; 3) The proposed hard voxel weighting strategy used for training effectively improves the segmentation accuracy; 4) The segmentation uncertainty obtained by our method has a high correlation to mis-segmentations, which has a potential to assist more informed decisions  in clinical practice.  Our code is available at https://github.com/HiLab-git/SepNet.

\end{abstract}


\end{frontmatter}



\section{Introduction}

Nasopharyngeal carcinoma (NPC) is a tumor arising from the epithelial cells that cover the surface and line the nasopharynx~\cite{wei2005nasopharyngeal}. It is one of leading form of cancer in the Arctic, China, Southeast Asia, and the Middle East/North Africa~\cite{chang2006enigmatic}. Radiation therapy is the most common treatment choice for NPC because it is mostly radio-sensitive~\cite{hunt2001treatment}. Radiation planning sets up the radiation dose distribution for tumor and ordinary organs, which is vital for the treatment. Oncologists would design radiotherapy plans to make sure the cancer cells receive enough radiation and prevent the damage of normal cells in Organs-at-Risk (OARs), such as the optical nerves and chiasma that are important for the vision of patients.
\par Medical images play a vital role in preoperative decision making, as they can assist radiologists to determine the OARs boundaries. Magnetic Resonance Imaging (MRI) and Computed Tomography (CT) are widely used non-invasive imaging methods for HAN. Delineating the boundaries of tens of OARs from HAN CT scans is one crucial step in radiotherapy treatment planning, and manual delineation is tedious and time consuming and likely to have high inter- and intra-observer variations. Automated segmentation of OARs can save radiologists time and has a potential to provide an accurate and reliable solution~\cite{zhu2019anatomynet}.

\par
\begin{figure*}
    \centering
    \includegraphics[width=0.85\linewidth]{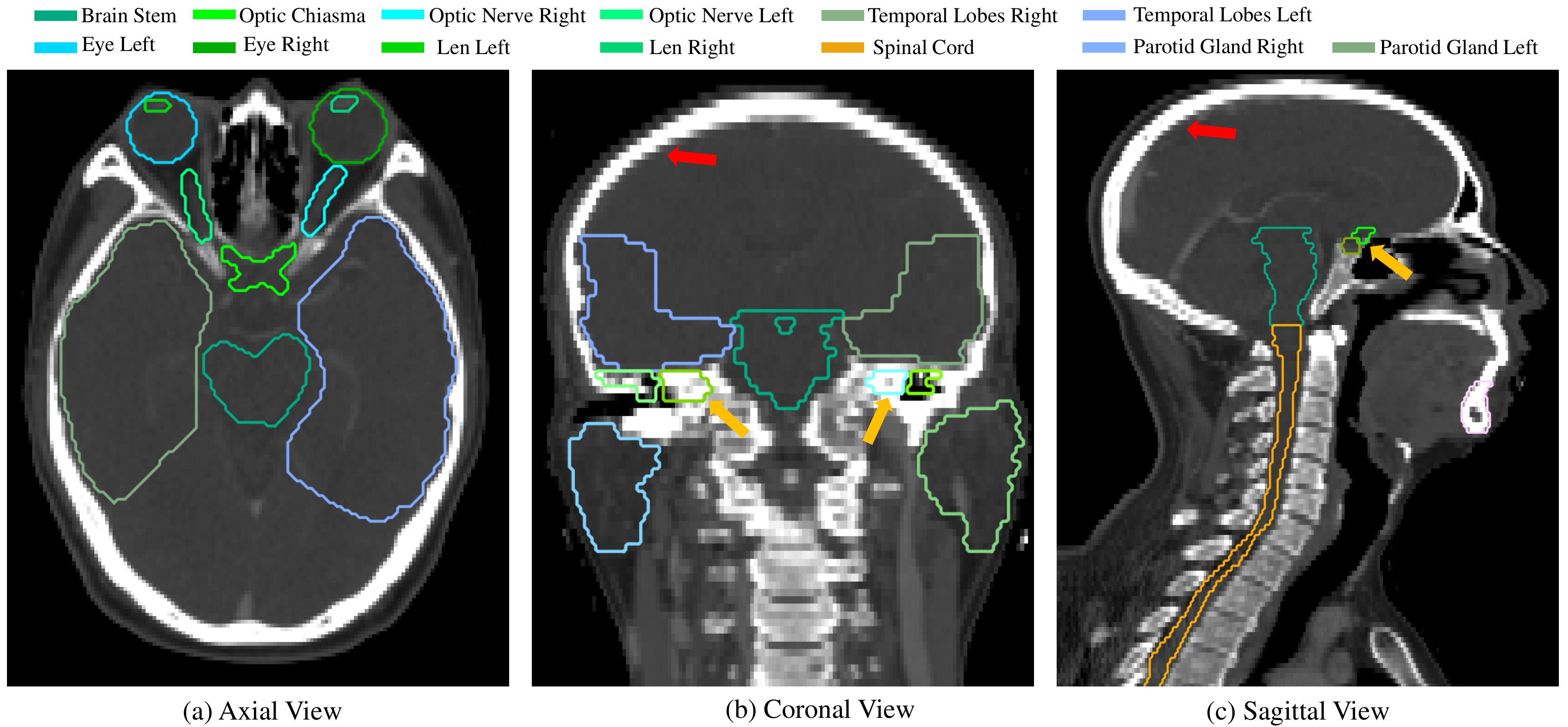}
    \caption{An example of HAN CT images in three orthogonal views. The OARs have imbalanced sizes with a low contrast with the surrounding tissues.  Note the large inter-slice spacing highlighted by red arrows and small organs occupying only few slices (e.g.), the optical nerves and the lens highlighted by yellow arrows.}
    \label{fig:organ_scale}
\end{figure*}

 However, the automatic segmentation of HAN OARs from CT images is still challenging because of several reasons. First, there is a severe imbalance between sizes of large and small OARs. As shown in Fig.~\ref{fig:organ_scale}, the smallest organs such as the lens are only a few voxels wide, while the parotid gland is over 250 times larger than the lens, making the automatic segmentation algorithm easily biased towards the large organs.
Nowadays, existing methods~\cite{zhu2019anatomynet,gao2019focusnet, wong20183d} mainly deal with the imbalance problem  by weighting small OARs in the loss function for training, which is a class-level weighting and ignores hard voxels in easy or large OARs. Second, due to the mechanism of CT imaging, soft tissues including OARs have a low contrast, such as the brain stem depicted in Fig.~\ref{fig:organ_scale}. The low contrast results in some uncertainty and challenges for accurate classification of voxels around the boundary of OARs. Most existing works~\cite{gao2019focusnet, wong20183d, nikolov2018deep} do not deal with this problem well as they use a single intensity window/level for intensity transform, where a naive linear function between a lower threshold and an upper threshold is used. Such a simple intensity transform cannot provide good visibility of multiple organs at the same time, which limits the accuracy of multi-organ segmentation. Thirdly, HAN CT images are usually acquired with high in-plane resolution while low through-plane resolution, resulting in the smaller slice number of tiny OARs, such as the optical nerves highlighted by yellow arrows in Fig.~\ref{fig:organ_scale}. Therefore, using standard 3D convolutions to treat $x$, $y$ and $z$ dimensions equally~\cite{zhu2019anatomynet, gao2019focusnet, wong20183d, nikolov2018deep}  may overlook the anisotropic spatial resolution and limit the accuracy of segmentation of these small organs. 
\par In real clinical applications, the uncertainty of predictions is important in OARs segmentation. For radiotherapy planning, radiologists  care about not only how accurate the segmentation results are, but also the extent to which the model is confident about its predictions. If the uncertainty is too high, then doctors need to focus on this part of area and may give refinements~\cite{shi2011multi}. In our case, the low contrast between  OARs and their surroundings leads the voxels around  boundaries to be segmented with less confidence~\cite{wang2019automatic}. The uncertainty information of these voxels can indicate regions that have potentially been mis-segmented, and could be used to guide human to refine the prediction results~\cite{lei2019deepigeos,wang2020miccai}. However, to the best of our knowledge, there have been no existing works exploring the uncertainty estimation in HAN OARs predictions and its relationship with mis-segmentation.

\par Though there have been some similar existing works on some components of the proposed method, our pipeline is elaborated for uncertainty-aware multi-organ segmentation from HAN CT images by considering the challenges associated with this specific task. In addition, several modules in our method are also different from their existing counterparts. First, to deal with low contrast of CT images, our Segmental Linear Functions (SLFs) transform the intensity of CT images to make multiple organs more distinguishable than existing methods based on a simple window width/level that often gives a better visibility of one organ while hiding the others. Second, existing general 2D and 3D segmentation networks can hardly deal with images with anisotropic spacing well. Our SepNet is a novel 2.5D network specially designed for dealing with clinic HAN CT scans with anisotropic spacing. Thirdly, existing hardness-aware loss functions often deal with class-level hardness, but our proposed attention to hard voxels (ATH) uses a voxel-level hardness strategy, which is more suitable to dealing with some hard regions despite that its corresponding class may be easy. Last but not least, uncertainty information is important for more informed treatment planning of NPC, but existing HAN CT segmentation methods do not provide the uncertainty information. We use an ensemble of models trained with different loss functions and intensity transforms to obtain robust results. Therefore, our proposed method is a novel pipeline for accurate segmentation of HAN OARs by better considering the image’s comprehensive properties including the low-contrast, anisotropic resolution, segmentation hardness, and it provides segmentation uncertainty for more informed treatment decision making. Our proposed method won the 3rd place in 17 teams in the HAN OAR segmentation task of StructSeg 2019$\footnote{https://structseg2019.grand-challenge.org/Home/}$ challenge . 

\section{Related Works}
\subsection{Traditional Methods for OAR Segmentation} 
In earlier studies, traditional segmentation methods such as atlas-\citep{daisne2013atlas}, RF-\cite{wang2017hierarchical}, contour-\cite{gorthi2009segmentation}, region-\cite{yu2009automated}, SVM-based~\cite{yang2014automated} methods, have achieved significant benefits compared with traditional manual segmentation.~\citet{zhou2006nasopharyngeal} proposed a SVM-based method to segment the cancer area from MRI. By drawing a hyperplane via kernel mapping, they are allowed to perform the task automatically without prior knowledge.~\citet{fitton2011semi} introduced a contour-based segmentation method, requiring the doctor to delineate a rough contour firstly. With underlying image characteristics, their “snake” algorithm will adjust the rough contour to achieve more accurate result. Applying region-grow algorithm,~\citet{lee2005segmentation} proposed a method requiring the user to select an initial seed within the tumor first, and then the seed grows to segment NPC lesions in MRI based on information obtained from contrast enhancement ratio.  
However, these methods are mainly based on low-level features and time consuming. 

\subsection{Deep Learning for OAR Segmentation}
After dramatic increase in popularity of deep learning methods in various computer vision tasks.~\citet{ibragimov2017segmentation} firstly proposed a CNN method at the end of 2016 to segment organs at HAN. They implemented a convolution neural network to extract features such as corners, end-points, and edges which contribute to classification and achieved higher accuracy than traditional methods. However, it requires further operations like Markov Random Fields algorithm to smooth the segmentation area and is patch-based, which would be time consuming for inference.
\par With the development of CNNs, a new architecture called Fully Convolutional Neural Network (FCN)~\cite{long2015fully} obtained large performance improvement for segmentation which also encouraged the emergence of U-Net~\cite{ronneberger2015u}.~\citet{men2017automatic} used deep dilated CNNs for segmentation of OARs, where the network has a larger receptive field. Other researchers~\cite{zhao2019automatic,fritscher2014automatic} presented a fully convolution network-based model with auxiliary paths strategy which allows the model to learn discriminative features to obtain a higher segmentation performance. To deal with the imbalance problem,~\citet{liang2019deep} used one CNN for OAR detection and then another CNN for fine segmentation. Based on 3D-UNet~\cite{cciccek20163d},~\citet{gao2019focusnet} proposed FocusNet by adding additional sub-networks for better segmentation of small organs. 
\par Loss functions also play an important role in the segmentation task. Liang et al.~\cite{liang2019deep} improved their loss function with weighted focal loss that performed well in the MICCAI 15 dataset~\cite{raudaschl2017evaluation}. Since the traditional Dice loss~\cite{milletari2016v} is unfavorable to small structures as a few voxels of misclassification can lead to a large decrease of the Dice score, and this sensitivity is irrelevant to the relative sizes among structures. Therefore, balancing by label frequencies is suboptimal for Dice loss. Exponential Logarithmic loss~\cite{wong20183d}, combining Focal loss~\cite{lin2017focal} with Dice loss, eased this issue without much adjustments of the network. However, it will reduce the accuracy of hard voxels in large or other easy regions.
\subsection{Networks for Anisotropic Spacing}
To deal with the anisotropic spacing in 3D image scans, previous networks can be divided into three major categories: First, directly applying 2D networks slice by slice to get the final results~\cite{gu2019net}, which is agnostic to large inter-slice spacing. However, due to the internal structural constraints, the 2D networks can hardly capture 3D information. Second, resampling the original images into the same spacing along three dimensions~\cite{chen2019lsrc, karimi2019reducing, heinrich2019closing} and processing them with standard 3D convolutions, which could merge the information among three dimensions effectively. But it will generate artifacts in interpolated slices and obtain reduced segmentation accuracy. Thirdly, applying 2.5D network~\cite{wang2017automatic, roth2015improving}, which firstly use 2D convolutions separately in three orthogonal views and fuse these multi-view information together in final stage, so that 3D contextual features can be used. However, previous 2.5D networks mainly treat each directions equally thus ignore the anisotropic spacing and organ scales among three dimensions. 
\subsection{Segmentation Uncertainty}
Despite the impressive performance of current methods has achieved, measuring how reliable the predictions are is also crucial to indicate potential mis-segmented regions or guide user interactions for refinement~\cite{prassni2010uncertainty, wang2018interactive, wang2020miccai}. There are two major types of predictive uncertainties for deep CNNs~\cite{kendall2017uncertainties}: $aleatoric$ uncertainty and $epistemic$ uncertainty. $Aleatoric$ uncertainty depends on noise or randomness in the input testing image, while $epistemic$ uncertainty, also known as model uncertainty, can be explained away given enough training data. A lot of works have investigated uncertainty estimation for deep neural networks~\cite{kendall2017uncertainties, lakshminarayanan2017simple, zhu2018bayesian} and they mainly focused on high-level image classification or regression tasks. Some recent works~\cite{roy2018inherent,nair2020exploring} investigated test-time dropout-based (epistemic) uncertainty for segmentation. ~\citet{wang2019aleatoric} and ~\citet{nair2020exploring} extensively investigate different kinds of uncertainties for CNN-based medical image segmentation, including not only epistemic but also aleatoric uncertainties. Even though uncertainty estimation is of great interest in HAN OARs segmentation, to the best of our knowledge, exiting works have not investigated this problem.   

\begin{figure*}
	\centering
	\includegraphics[width=1\linewidth]{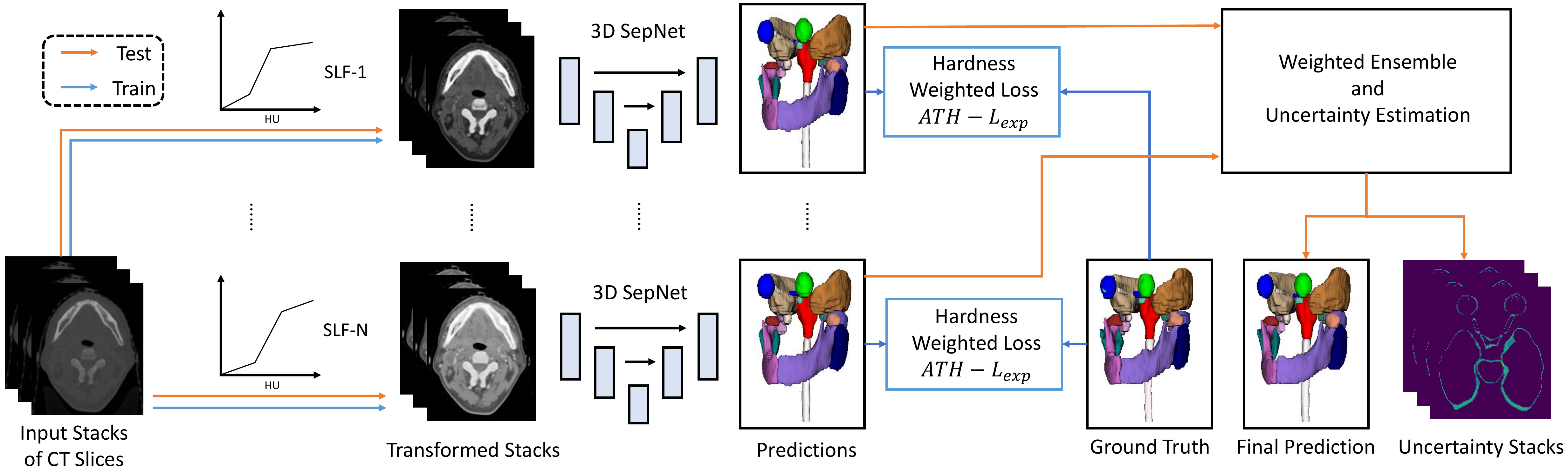}
	\caption{Overview of our proposed framework for accurate OARs segmentation. We first use different Segmental Linear Functions (SLFs) to transform the intensity of an input image respectively, which obtains good visibility of different OARs. Then a novel network 3D-SepNet that leverages intra-slice and inter-slice convolutions is proposed to segment OARs from  images with large inter-slice spacing, and we propose a novel hardness weighting strategy for training.  Finally, an ensemble of networks related to different SLFs and loss functions obtains the final segmentation and uncertainty estimation simultaneously. }
	\label{fig:summary}
\end{figure*}
\section{Methods}
Our proposed framework for HAN OARs segmentation is shown in Fig.~\ref{fig:summary}. It consists of four main components. %Though there have been some similar existing works on some components of the proposed method, our pipeline is elaborated for uncertainty-aware multi-organ segmentation from Head and Neck CT images by considering the challenges associated with this specific task. In addition, several modules in our method are also different from their existing counterparts. 
First, we propose to use  Segmental Linear Functions (SLF) to obtain multiple intensity-transformed copies of an input image to get better contrasts of different OARs. Second, for each copy of the intensity-transformed image, we employ a novel network 3D-SepNet combining intra-slice and inter-slice convolutions to deal with the large inter-slice spacing. Thirdly, to train our 3D-SepNet we propose a novel hard voxel weighting strategy that pays  more attention to small organs and hard voxels in large/easy organs, and it can be combined with existing loss functions. Finally, we ensemble several models trained with different SLFs and loss functions by a weighted fusion to get final segmentation results, which simultaneously obtains uncertainty estimation of the segmentation, as show in Fig.~\ref{fig:summary}.

\subsection{Intensity Transform with Segmental Linear Function}
\begin{figure}
	\centering\includegraphics[width=1\linewidth]{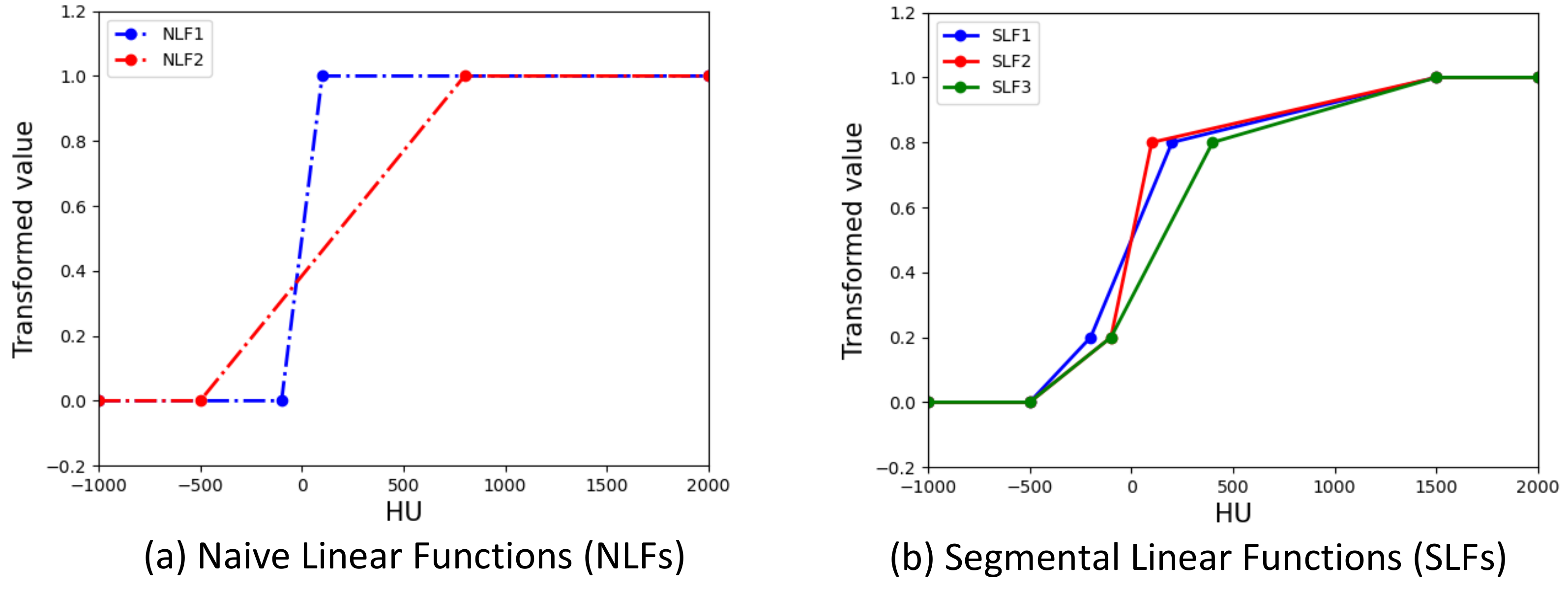}
	\caption{Different intensity transform functions for CT image preprocessing. Standard  window width/level-based intensity transforms use NLFs shown in (a), which employs a single linear function between the lower and the upper thresholds. Our proposed SLFs in (b) transform the HU values to $K$ linear sections to obtain better visibility of multiple OARs. 
	}
	\label{fig:slf}
\end{figure}
HU values of soft issues such as the brain stem, parotid and temporal lobes are very close, and largely different from those of bones in CT images. Previous studies~\cite{zhu2019anatomynet,nikolov2018deep,wong20183d} used simple window width/level-based intensity transform for preprocessing, i.e., a lower threshold and a upper threshold with a linear function between them, which is shown in Fig.~\ref{fig:slf}(a) and referred to as Naive Linear Functions (NLFs) in this paper. Processing CT images with an NLF is sub-optimal as it cannot obtain good visibility for soft tissues and bones at the same time, which may limit the segmentation accuracy. To address this issue and inspired by the fact that radiologists would use different window widths/levels to better differentiate OARs with diverse HU values, we propose to use Segmental Linear Functions (SLFs) to transform the CT images. More specifically, assume increasing numbers $[x_1, x_2, ..., x_K]$ in [0,1] and corresponding $K$ HU values $[h_1, h_2, ..., h_K]$ in $[h_{min}, h_{max}]$, where $h_{min}$ and $h_{max}$ represent the minimal and maximal HU value of CT images, respectively. Assume the original HU value is $h$, and the transformed intensity $x$ is represented as:
\begin{equation}
    x = \left\{\begin{matrix}
     0, & h\leq h_1\\ 
     h_i+\frac{h-h_i}{h_{i+1}-h_i}, & h_i < h \leq h_{i+1}, \ i\in[1,K]
    \\
     1 , & h>h_K
\end{matrix}\right.
\end{equation}
\par In this work, we set  $K=4$, $[x_1,x_2,x_3,x_4]$=[0, 0.2, 0.8, 1.0], and use three different SLFs: SLF1, SLF2, SLF3, with $[h_1,h_2,h_3,h_4]$ set as [-500, -200, 200, 1500], [-500, -100, 100, 1500] and [-500, -100, 400, 1500], respectively.  We also compared them with two types of NLFs: NLF1, NLF2, with corresponding $[h_1,h_2]$ set as [-100, 100] to focus on soft tissues and [-500, 800] for large window width, respectively, as shown in Fig.~\ref{fig:slf}(b). 

\begin{figure}
\centering\includegraphics[width=\linewidth]{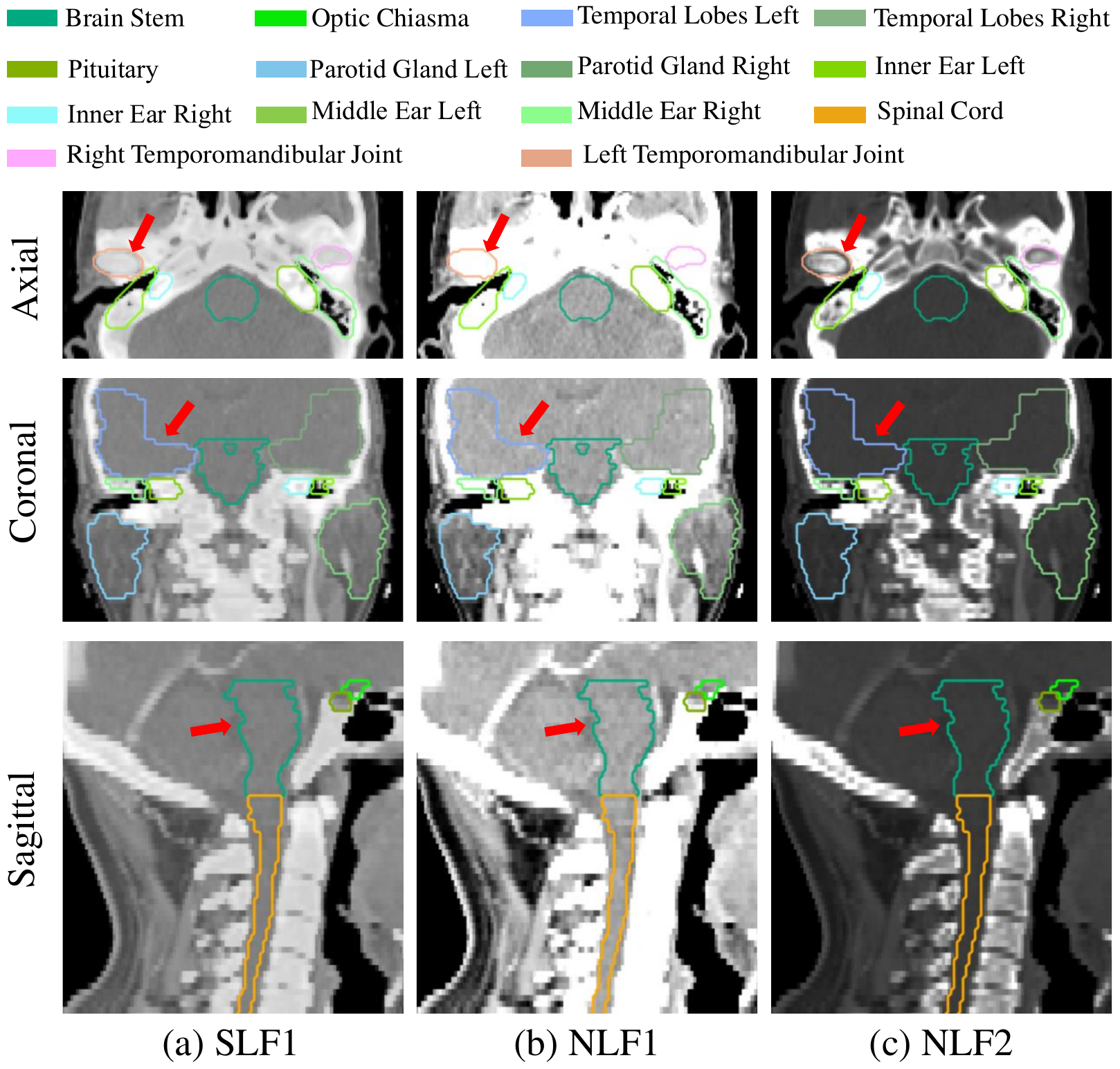}
\caption{Visualization of a CT image with different intensity transform functions. The sub-figures from left to right are transformed CT images based on: (a) our SLF1 that gives a better visualization to both bone and soft tissues, (b) NLF1 that highlights soft tissues and (c) NLF2 that preserves the internal structure and boundary shape of bone tissues while reduce the visibility of soft tissues.}
\label{fig:multi_transform_comparision}
\end{figure}
\begin{figure*}
	\centering\includegraphics[width=1\linewidth]{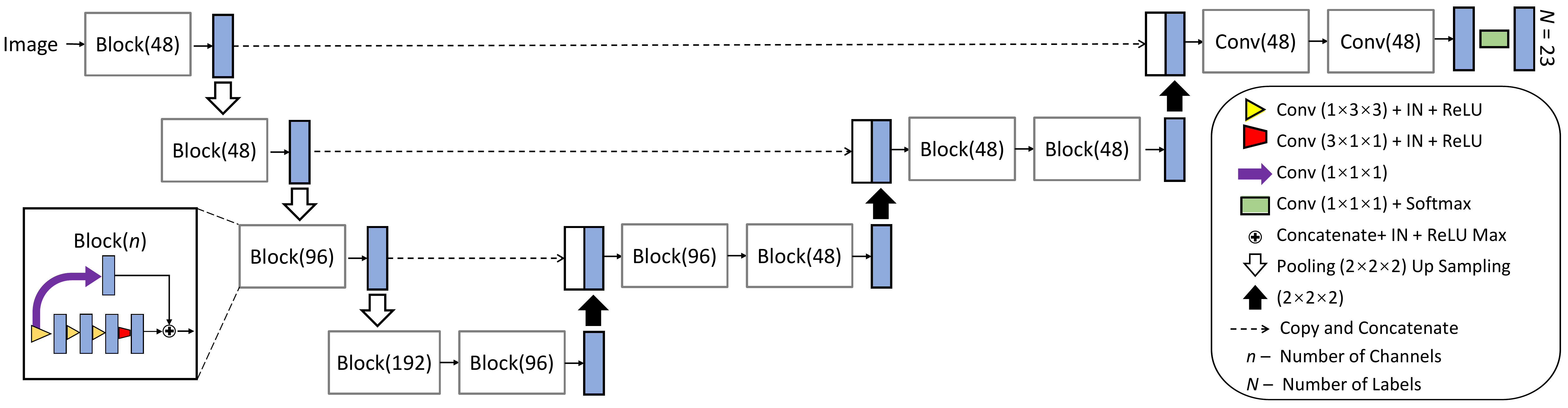}
	\caption{Our proposed network 3D-SepNet for  segmentation of OARs from images with large inter-slice spacing. Blue and white boxes indicate feature maps. Block($n$) represents a convolutional block with $n$ channels, where three intra-slice  layers (yellow arrows) are followed by one inter-slice convolutional layer (red arrow) and a skip connection with one $1\times1\times1$ convolutional layer is used for better convergence.}
	\label{fig:network architecture}
\end{figure*}
\par Fig.~\ref{fig:multi_transform_comparision} shows a visual comparison of different intensity transforms (i.e., SLF1, NLF1 and NLF2) applied to HAN CT scans. From axial view, it could be observed that NLF1 suppresses the bones and improves the visibility of soft tissues, while NLF2 improves the visibility of bones but makes soft tissues hardly distinguishable. Each of these intensity transforms makes it hard to segment multiple OARs including bones (e.g., mandible) and soft tissues at the same time. In contrast, 
our SLF1  obtains high visibility for both soft tissue and bones, which is beneficial for the segmentation of multiple OARs with complex intensity distributions.  

\subsection{Network Architecture}
Due to the large inter-slice spacing and sharp boundary between bones and nearby soft tissues, upsampling the images along the $z$ axis to obtain a high 3D isotropic resolution will produce plenty of artifact on the interpolated slices and mislead the segmentation results. Therefore, we directly use the stacks of 2D slices for segmentation. As 2D networks ignore the correlation between adjacent slices and standard 3D networks with isotropic 3D convolutions (e.g., with $3\times3\times3$ kernels) lead to imbalanced physical receptive field (in terms of $mm$ rather than voxels) along each axis with different voxel spacings, we propose a 2.5D network combining intra-slice convolutions and inter-slice convolutions to deal with this problem. 

As shown in Fig.~\ref{fig:network architecture}, our proposed network (3D SepNet) is based on the backbone of 3D UNet~\cite{cciccek20163d} with 12 convolutional blocks in an encoder-decoder structure. Since the inter-slice and intra-slice voxel spacings of our experimental images are around 3~mm and 1~mm respectively, small organs like optical nerves/chiasma would only cross few slices. Therefore, applying standard 3D convolutions would blur their boundaries and be harmful to accuracy. We proposed to use spatially separable convolutions that separate a standard 3D convolution with a $3\times3\times3$ kernel into an intra-slice convolution with a $1\times3\times3$ kernel and an inter-slice convolution with a $3\times1\times1$ kernel. Considering the large inter-slice spacing, we use one inter-slice convolution after each three intra-slice convolutions, as shown in Fig.~\ref{fig:network architecture}. The spatially separable convolution is implemented in Pytorch by simply specifying the kernel size. We use Instance Normalization (IN)~\cite{ulyanov2016instance} and Rectified Linear Units (ReLU)~\cite{nair2010rectified} after each convolutional layer. A skip connection with a $1\times1\times1$ convolutional layer is used in each block for better convergence. The number of channels ($n$) is doubled after each max pooling in the encoder. We concatenate feature maps from the encoding path with the corresponding feature maps in the decoding path for better performance. A final layer of $1\times1\times1$ convolution with the softmax function provides the segmentation probabilities.

\subsection{Attention to Hard voxels}
In our task, the ratio between the background and the smallest organ can reach around $10^{5}:1$, which makes the loss function values dominated by large numbers of easy background voxels. To solve this problem, we first use the exponential logarithmic loss~\cite{wong20183d} $(L_{Exp})$, which balances the labels not only by their relative sizes but also by their segmentation difficulties:
\begin{equation}
    L_{Exp} = \omega_{DSC}L_{DSC} + \omega_{Cross}L_{Cross}
\end{equation}
with $\omega_{DSC}$ and $\omega_{Cross}$ are the weights of the exponential logarithmic DSC loss ($L_{DSC}$) and the exponential cross-entropy ($L_{Cross}$), respectively.
\begin{equation}
    L_{DSC} = \bm{E}[(-ln(DSC_c))^{\gamma_{DSC}}]
\end{equation}
\begin{equation}
    DSC_c = \frac{2[\sum_xg_c(\bm{x})p_c(\bm{x})]+\epsilon}{ \{\sum_x[g_c(\bm{x})+p_c(\bm{x})] \}+\epsilon}
\end{equation}
\begin{equation}
    L_{Cross} = \bm{E}\{w_c\{-ln[p_c(\bm{x})]\}^{\gamma_{Cross}}\}
\end{equation}
where $\bm{x}$ is a voxel and $c$ is a class. $p_c(\bm{x})$ is the predicted probability of being class $c$ for voxel $\bm{x}$, and $g_c(\bm{x})$ is the corresponding ground truth label. $\bm{E[\cdot]}$ is the mean value with respect to $c$ and $\bm{x}$ in $L_{DSC}$ and $L_{Cross}$, respectively.  $\epsilon=1$ is the pseudocount for additive smoothing to handle missing labels in training samples. $w_c = ((\sum_k f_k)/f_c)^{0.5}$ is the class-level weight for reducing the influence of more frequently seen classes, where $f_k$ is the frequency of class $k$. We set $\omega_{DSC} , \omega_{Cross}, \gamma_{DSC}, \gamma_{Cross}=1$ in our experiments according to~\cite{wong20183d}.
\par Unlike $L_{Cross}$, $L_{DSC}$ weights voxels in one certain label mainly considering its segmentation difficulty and the formulation of $L_{DSC}$ can be differentiated yielding the gradient :
\begin{equation}
\begin{aligned}
        \frac{\partial L_{DSC}}{\partial p_c(\bm{x})}&=
        \frac{\partial L_{DSC}}{\partial DSC_c}\frac{\partial DSC_c}{\partial p_c(\bm{x})}\\
        &=-\frac{1}{DSC_c}\frac{2\{g_{c}(\bm{x})  \sum_{\bm{x}}[g_{c}(\bm{x})+p_c(\bm{x})] -\sum_{\bm{x}}p_c(\bm{x})g_{c}(\bm{x})\}}{ \{\sum_{\bm{x}}[g_{c}(\bm{x})+p_c(\bm{x})] \}^2}
\end{aligned}
\end{equation}
\par It is easy to observe that for a hard voxel $\bm{x}$ in easy class $c$, with  $p_c(\bm{x})$ far away from $g_{c}(\bm{x})$, the absolute value of gradient $\frac{\partial L_{DSC}}{\partial DSC_c}$ is restrained compared with a voxel $\bm{x}$ in hard region because the value $\frac{\partial L_{DSC}}{\partial DSC_c}$ is constrained around 1 with $DSC_c\approx1$. Therefore, it may limit the performance of CNNs on hard voxels, most of which lie on the the boundary between different objects, in large or easy regions, and it will be helpful to force the network to focus on the hard voxels to balance this contradiction. More formally, we propose to multiply the prediction result by a weighting function $w_c(\bm{x})$ before sending it into $L_{DSC}$, with a tunable attention parameter $\alpha>0$:
\begin{equation}
        w_c(\bm{x}) = e^{\frac{p_c(\bm{x})-g_c(\bm{x})}{\alpha}}
\end{equation}
\begin{equation}
    p^w_c(\bm{x}) = p_c(\bm{x})w_c(\bm{x})
\end{equation}
where $p^w_c(\bm{x})$ is the weighted prediction. A higher value of $p^w_c(\bm{x})$ represents a higher possibility of voxel $\bm{x}$ belonging to class $c$. 
\begin{figure}
	\centering
	\subfigure[$g_c=1$]{
    \begin{minipage}{0.49\linewidth}
    \centering
    \includegraphics[width=1\linewidth]{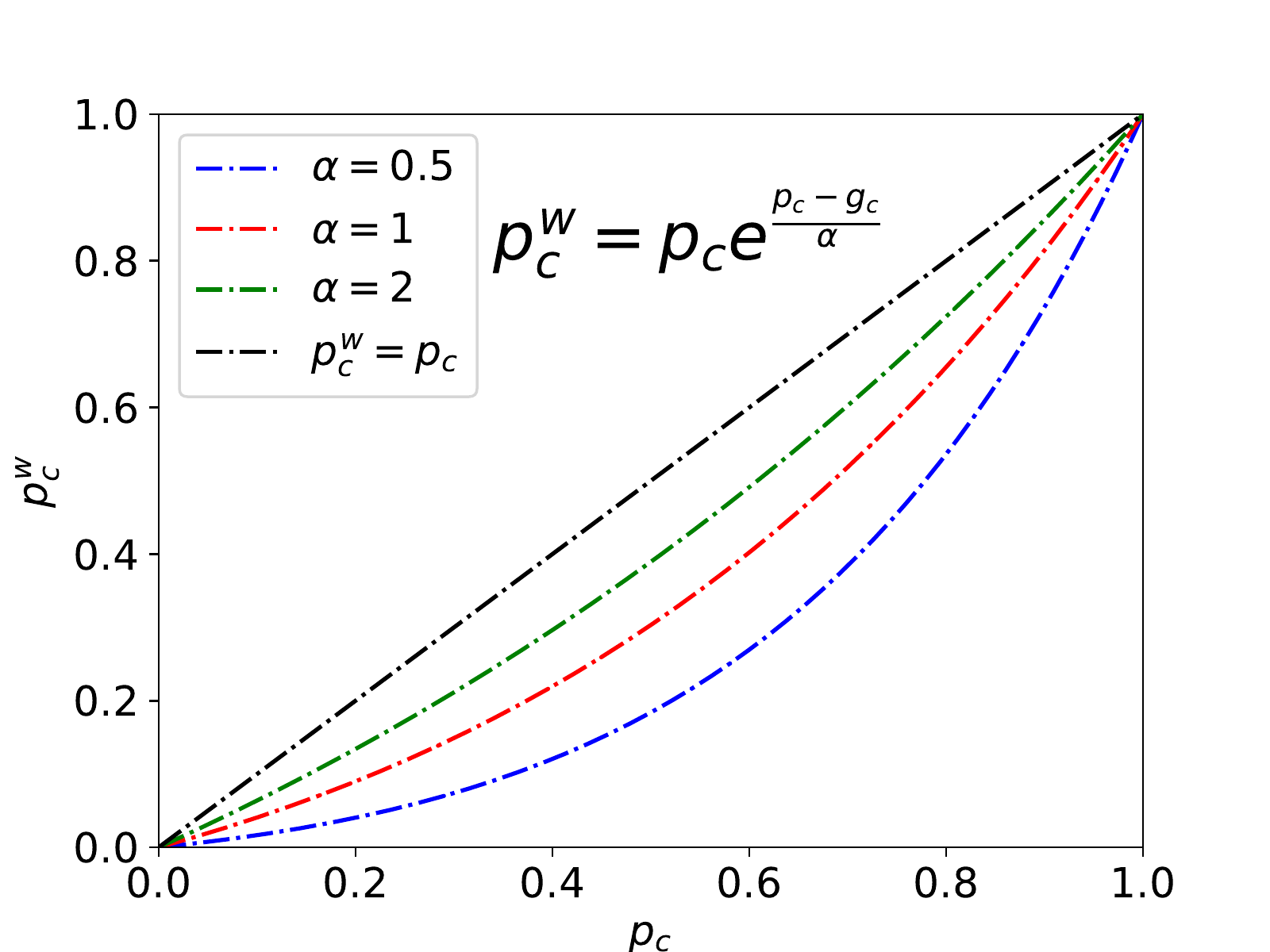}
    \end{minipage}%
    }%
    \subfigure[$g_c=0$]{
    \begin{minipage}{0.49\linewidth}
    \centering
    \includegraphics[width=1\linewidth]{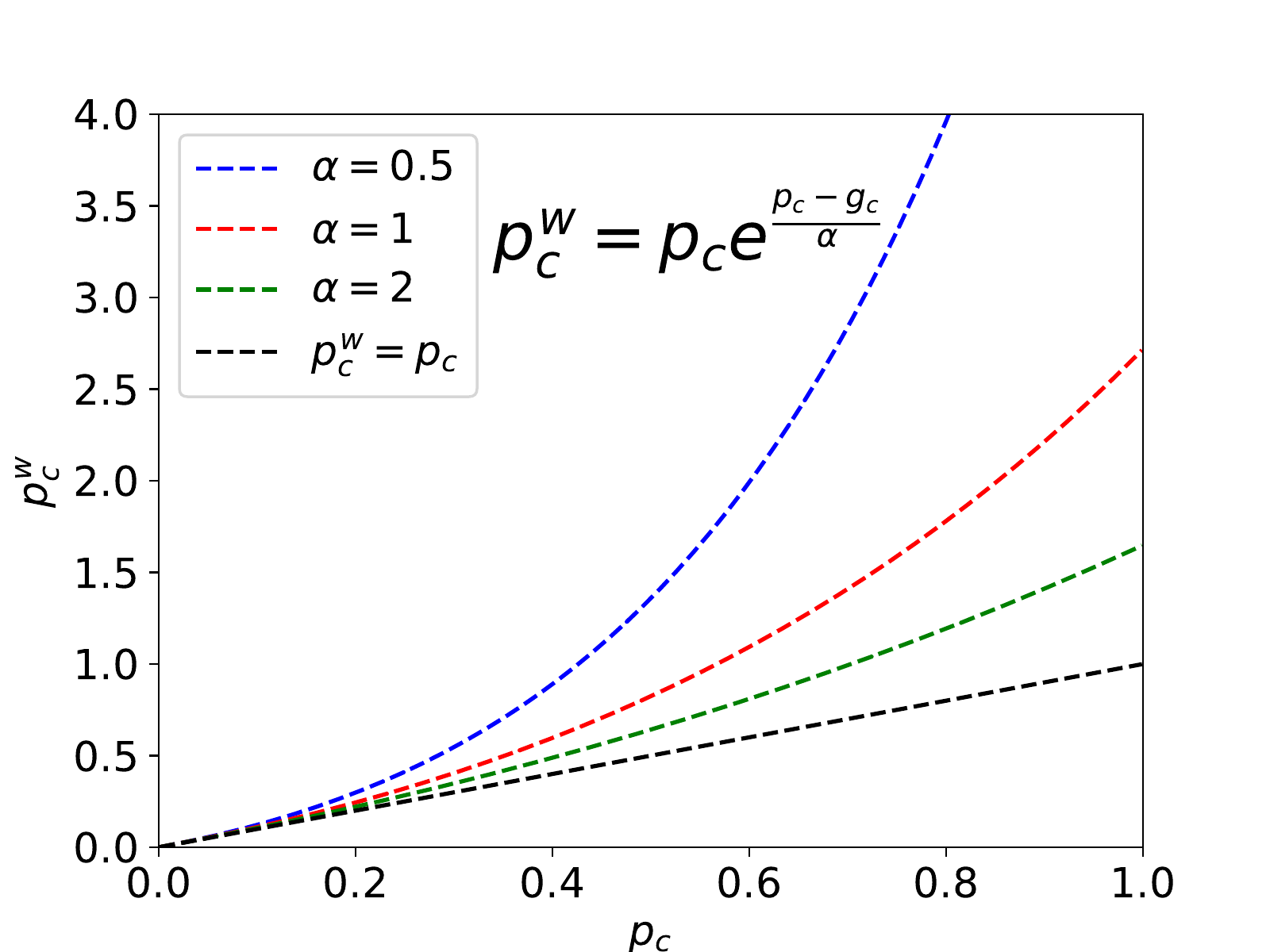}
    \end{minipage}%
    }%
	\caption{Weighted prediction $p_c^w$ with different $\alpha$ and ground truth label $g_c$ for a voxel. $p_c^w$ is lower than $p_c$ for $g_c=1$ and higher for $g_c=0$, meaning the weighted prediction is further away from the ground truth than $g_c$, which leads a less accurate prediction to have a larger impact on the backpropagation.}
	\label{fig:attention weight map}
\end{figure}

Fig.~\ref{fig:attention weight map} shows the effect of weighting function $w_c(\bm{x})$ with different values of $\alpha$ for $g_c = 1$ and $g_c=0$, respectively. It can be observed that  $p^w_c(\bm{x})$ is lower than $p_c(\bm{x})$ for $g_c(\bm{x})=1$ and higher for $g_c(\bm{x})=0$, meaning the weighted prediction is further away from the ground truth than the original prediction. As a result, the weighted harder region will have a larger impact on the backpropagation as they have larger gradient values than accurately predicted voxels. Generally, it can get more room for improvement and have the network focus more on hard voxels. We refer to this weighting strategy as attention to hard voxels ($ATH$). It should be noticed that $ATH$ can be combined with a standard loss function in the training progress, and we combine it with $L_{Exp}$ due to its better performance than standard Dice loss and cross entropy loss. Our loss function with $ATH$ is  named as $ATH-L_{Exp}$.

\label{section:3.3}

\subsection{Model Ensemble and Uncertainty Estimation}

To achieve robust segmentation and obtain segmentation uncertainty at the same time, we use an ensemble of models trained with different SLFs and ATHs based on different $\alpha$, as shown in Fig.~\ref{fig:summary}. Since these models perform differently on different organs, we implement the ensemble for each class respectively and use a class-wise weighted average for ensemble. Specifically, for class $c$, a model with a higher performance on $c$  is assigned with a higher weight. For a model obtaining the $i$th highest DSC for class $c$ of the validation set, its predicted probability map of class $c$ and the corresponding weight for a test image are denoted as $P^i_{c}$ and $w^i_{c}$ respectively. We set $w^i_{c}$ to  5, 4, 3, 1, 1, 1 for $i$=1, 2, ...,6 respectively. The final probability map for class  $c$ of a test image is:
 \begin{equation}
     \hat{P}_c=\frac{\sum_i w^i_c P^i_c}{\sum_i w^i_c}
 \end{equation}
 
\par Based on the predictions of multiple models, it is straightforward to obtain the segmentation uncertainty that is typically estimated by measuring the diversity of multiple predictions for a given image~\cite{kendall2017uncertainties}. Supposing $X$ denotes a test image  and $Y$  the predicted label of $X$. Using the variance and entropy of the distribution $p(Y|X)$ are two common methods for  uncertainty estimation. Similarly to~\cite{lakshminarayanan2017simple}, it is a promising choice to estimate uncertainty by ensemble method, using multiple independently learned models. We use the entropy to represent the voxel-wise uncertainty:
 \begin{equation}
     H(Y|X)=-\int p(y|X)\text{ln}(p(y|X))dy
 \end{equation}
\par Suppose $Y(\bm{x})$ denotes the predicted label for voxel $\bm{x}$. With predictions from $N$ models, a set of values  $\mathbb{Y}=\{y_1(\bm{x}), y_2(\bm{x}), ..., y_N(\bm{x})\}$ can be obtained. Therefore, the voxel-wise uncertainty can be approximated as:
\begin{equation}\label{eq:voxel_uncertain}
    H(Y|X) \approx -\sum_{m=1}^{M}\hat{p}_m(\bm{x})\text{ln}(\hat{p}_m(\bm{x}))
\end{equation}
where $\hat{p}_m(\bm{x})$ is the frequency of the $m$th unique value in $\mathbb{Y}$.
\par We also estimate the structure-wise uncertainty for each OAR by calculating the Volume Variation Coefficient (VVC). Let $V^i=\{v_1^i, v_2^i, ..., v_N^i\}$ denote the set of volumes of OAR $i$ obtained by $N$ models, and $\mu_{V^i}$ and $\sigma_{V^i}$ denote the mean value and standard deviation of $V^i$ respectively. We use the VVC to estimate the structure-wise uncertainty for OAR $i$:
\begin{equation}
    {VVC}^i=\frac{\mu_{V^i}}{\sigma_{V^i}}
\end{equation}

\section{Experiments and Results}
\begin{table*}
	\caption{DSC (mean$\pm$std \%) and ASSD (mean$\pm$std mm) evaluation of 3D HAN OARs segmentation in StructSeg 19 task1, with 3D-SepNet, 3 different intensity transform methods and 3 different loss functions. The best result is in bold font.}
	\centering
	\scalebox{0.75}{
		\centering
		\begin{tabular}{c |c c c c c c| c c c c c c}
			\hline
			\textbf{Metric}
			&
			\multicolumn{5}{c}{DSC (mean$\pm$std $\%$)$\uparrow$} &&&
			\multicolumn{5}{c}{95$\%$HD (mean$\pm$std mm)$\downarrow$}
			\\
			\cline{1-13}
			\textbf{Transform}& 
			NLF1 & NLF2 & \multicolumn{3}{|c}{\textbf{SLF1 (ours)}}& &&
			NLF1& NLF2 & \multicolumn{3}{|c}{\textbf{SLF1 (ours)}}
			\\
			\cline{1-13}
			\textbf{Loss}&
			\multicolumn{3}{c|}{$\bm{ATH-L_{Exp}}$ \textbf{(ours)}}& 
			% $ATH-L_{Exp}$& 
			\multicolumn{1}{c|}{$L_{Exp}$} & 
			DSC Loss&&&
			\multicolumn{3}{c|}{$\bm{ATH-L_{Exp}}$ \textbf{(ours)}}& 
			\multicolumn{1}{c|}{$L_{Exp}$} & 
			DSC Loss
			\\ \hline
			Brain Stem & 88.0$\pm$2.0 & 88.9$\pm$1.6 & $\bm{89.7{\pm}1.7}$ & 81.5$\pm$2.4 & 88.0$\pm$2.0 &  &  & 3.11$\pm$0.18 & 2.82$\pm$0.26 & $\bm{2.79{\pm}0.27}$ & 3.93$\pm$0.41 & 3.13$\pm$0.26 \\
			Eye L & $\bm{89.2{\pm}3.0}$ & 88.3$\pm$4.1 & 88.6$\pm$3.4 & 87.2$\pm$4.8 & 86.0$\pm$5.0 &  &  & $\bm{2.38{\pm}0.32}$ & 2.44$\pm$0.26 & 2.47$\pm$0.31 & 2.53$\pm$0.36 & 2.56$\pm$0.38 \\ 
			Eye R & 87.9$\pm$4.6 & $\bm{88.0{\pm}2.8}$ & 87.3$\pm$2.5 & 86.1$\pm$3.3 & 86.2$\pm$3.0 &  &  & $\bm{2.44{\pm}0.60}$ & 2.45$\pm$0.20 & 2.56$\pm$0.20 & 2.46$\pm$0.22 & 2.59$\pm$0.23 \\ 
			Lens L & 80.5$\pm$6.5 & $\bm{81.9{\pm}6.2}$ & 81.5$\pm$6.9 & 75.0$\pm$8.2 & 80.4$\pm$6.5 &  &  & 2.42$\pm$0.39 & $\bm{2.37{\pm}0.38}$ & $\bm{2.37{\pm}0.38}$ & 2.52$\pm$0.40 &  2.42$\pm$0.42 \\ 
			Lens R & 75.7$\pm$9.8 & $\bm{77.2{\pm}10.9}$ & 75.4$\pm$13.0 & 74.5$\pm$9.1 & 77.8$\pm$9.9 &  &  & 2.51$\pm$0.44 & $\bm{2.38{\pm}0.32}$ & 2.47$\pm$0.45 & 2.59$\pm$0.52 & 2.39$\pm$0.37 \\ 
			Opt Nerve L & 68.5$\pm$12.1 & $\bm{69.8{\pm}10.3}$ & 68.0$\pm$10.7 & 67.2$\pm$12.4 & 69.1$\pm$12.1 &  &  & 2.77$\pm$0.64 & $\bm{2.74{\pm}0.67}$ & 2.90$\pm$0.78 & 2.83$\pm$0.63 & 2.29$\pm$0.88 \\ 
			Opt Nerve R & 64.2$\pm$13.9 & 63.9$\pm$11.3 & $\bm{66.3{\pm}11.7}$ & 61.4$\pm$13.8 & 64.9$\pm$12. &  &  & 3.05$\pm$1.04 & 3.04$\pm$0.84 & $\bm{2.93{\pm}0.77}$ & 3.18$\pm$1.03 & 3.07$\pm$1.07 \\ 
			Opt Chiasma & 54.9$\pm$8.5 & 53.5$\pm$11.2 & $\bm{56.6{\pm}8.7}$ & 49.1$\pm$10.4 & 55.5$\pm$13.6 &  &  & 4.07$\pm$1.33 & $\bm{3.72{\pm}0.79}$ & 3.83$\pm$0.87 & 4.71$\pm$1.49 & 3.70$\pm$0.55 \\ 
			Temporal Lobes L & 86.2$\pm$2.6 & $\bm{86.3{\pm}3.8}$ & 86.2$\pm$3.3 & 84.3$\pm$3.1 & 85.8$\pm$3.0 &  &  & 4.39$\pm$1.10 & 4.43$\pm$1.02 & $\bm{4.31{\pm}0.85}$ & 4.66$\pm$0.86 & 4.58$\pm$0.88 \\ 
			Temporal Lobes R & 83.5$\pm$4.5 & $\bm{86.9{\pm}3.1}$ & 86.4$\pm$2.8 & 83.6$\pm$2.9 & 86.8$\pm$2.4 &  &  & 5.43$\pm$1.69 & 4.52$\pm$0.93 & $\bm{4.38{\pm}0.77}$ & 4.88$\pm$0.75 & 4.33$\pm$0.70 \\
			Pituitary & 64.6$\pm$22.6 & 62.8$\pm$19.1 & $\bm{66.1{\pm}16.3}$ & 62.2$\pm$21.4 & 65.3$\pm$23.6 &  &  & 3.15$\pm$1.31 & 3.01$\pm$0.96 & $\bm{2.92{\pm}1.01}$ & 3.41$\pm$1.49 & 3.21$\pm$1.46 \\ 
			Parotid Gland L & 84.9$\pm$3.4 & 85.1$\pm$3.3 & $\bm{85.7{\pm}3.0}$ & 79.4$\pm$3.1 & 84.5$\pm$3.5 &  &  & 3.91$\pm$0.52 & 3.82$\pm$0.66 & $\bm{3.71{\pm}0.64}$ & 4.68$\pm$0.77 & 3.98$\pm$0.73 \\ 
			Parotid Gland R & 85.2$\pm$2.8 & 86.1$\pm$2.2 & $\bm{87.3{\pm}3.0}$ & 78.9$\pm$3.9 & 85.1$\pm$2.0 &  &  & 3.70$\pm$0.58 & 3.60$\pm$0.49 & $\bm{3.49{\pm}0.66}$ & 4.68$\pm$1.09 & 3.97$\pm$0.90 \\ 
			Inner Ear L & 79.4$\pm$5.4 & 78.9$\pm$5.2 & $\bm{80.3{\pm}6.9}$ & 74.0$\pm$6.4 & 79.1$\pm$7.0 &  &  & $\bm{2.61{\pm}0.47}$ & 2.85$\pm$0.76 & 2.85$\pm$0.62 & 3.28$\pm$0.87 & 2.74$\pm$0.92 \\ 
			Inner Ear R & 79.6$\pm$7.4 & 78.0$\pm$6.7 & $\bm{80.1{\pm}6.8}$ & 73.2$\pm$8.0 & 78.3$\pm$8.1 &  &  & 2.77$\pm$0.46 & 2.95$\pm$0.58 & $\bm{2.77{\pm}0.45}$ & 3.20$\pm$0.63 & 2.85$\pm$0.53 \\ 
			Mid Ear L & 81.8$\pm$8.4 & 82.0$\pm$9.8 & 82.6$\pm$5.7 & 78.6$\pm$10.9 & $\bm{83.5{\pm}5.4}$ &  &  & 3.77$\pm$1.58 & 4.65$\pm$1.68 & $\bm{3.33{\pm}0.87}$ & 4.34$\pm$2.38 & 3.57$\pm$1.40 \\ 
			Mid Ear R & 77.2$\pm$11.0 & 78.4$\pm$10.4 & 78.3$\pm$11.9 & 76.3$\pm$10.5 & $\bm{78.5{\pm}9.9}$ &  &  & 3.65$\pm$1.00 & 3.37$\pm$0.91 & 3.54$\pm$1.24 & 3.77$\pm$1.09 & $\bm{3.47{\pm}0.81}$ \\ 
			TM Joint L & $\bm{77.2{\pm}7.9}$ & 75.7$\pm$7.4 & 75.7$\pm$9.5 & 74.3$\pm$8.9 & 76.7$\pm$8.7 &  &  & $\bm{2.74{\pm}0.58}$ & 2.97$\pm$0.40 & 2.85$\pm$0.70 & 2.78$\pm$0.45 & 2.80$\pm$0.65 \\ 
			TM Joint R & 78.1$\pm$7.0 & 75.8$\pm$6.3 & 77.2$\pm$7.8 & 72.5$\pm$8.0 & $\bm{78.7{\pm}7.5}$ &  &  & $\bm{2.47{\pm}0.42}$ & 2.72$\pm$0.52 & 2.63$\pm$0.51 & 2.94$\pm$0.42 & 2.54$\pm$0.46 \\ 
			Spinal Cord & $\bm{83.1{\pm}2.5}$ & 82.1$\pm$4.7 & 83.0$\pm$4.0 & 73.3$\pm$6.1 & 82.7$\pm$4.0 &  &  & 2.89$\pm$0.24 & 2.84$\pm$0.32 & $\bm{2.78{\pm}0.33}$ & 3.42$\pm$0.23 & 2.84$\pm$0.32 \\
			Mandible L & 90.4$\pm$3.6 & 89.3$\pm$3.8 & $\bm{91.1{\pm}2.9}$ & 85.9$\pm$4.0 & 89.5$\pm$3.0 &  &  & $\bm{2.68{\pm}0.63}$ & 3.16$\pm$0.67 & 2.81$\pm$0.45 & 3.71$\pm$0.96 & 3.31$\pm$0.68 \\ 
			Mandible R & $\bm{91.8{\pm}2.8}$ & 90.5$\pm$3.0 & 91.7$\pm$1.5 & 88.3$\pm$2.6 & 91.6$\pm$2.1 &  &  & $\bm{2.55{\pm}0.53}$ & 2.72$\pm$0.43 & 2.70$\pm$0.40 & 3.07$\pm$0.57 & 3.67$\pm$0.50 \\ \hline
			\textbf{Average} & 79.6$\pm$2.4 & 79.5$\pm$2.3 & $\bm{80.2{\pm}2.3}$ & 75.8$\pm$1.8 & 79.7$\pm$2.3 &  &  & 3.16$\pm$0.29 & 3.16$\pm$0.29 & $\bm{3.06{\pm}0.24}$ & 3.53$\pm$0.27 & 3.17$\pm$0.20 
			\\  \hline
		\end{tabular}
	}
	\label{tab:trans_and_loss_compare}
\end{table*}
Our proposed methods are evaluated on two segmentation tasks: (1) StructSeg 2019 challenge training dataset consisting of CT scans from 50 NPC patients where 22 OARs are to be segmented. (2) A mixed HAN CT dataset containing 165 volumes from three sources: 50 volumes from StructSeg 2019, 48 volumes from MICCAI 2015 Head and Neck challenge~\cite{raudaschl2017evaluation} and 67 volumes collected locally with NPC before radiotherapy treatment, where we segment 7 OARs that are annotated in all of these three sources. For comparison, we implemented 3D-UNet~\cite{cciccek20163d} with three versions: original one, SE block~\cite{hu2018squeeze} added and residual connection~\cite{he2016deep} added, where we modified them by starting with 48 base channels and applying IN~\cite{ulyanov2016instance} for normalization. These networks and our proposed 3D-SepNet were implemented in PyTorch, trained on two NVIDIA 1080TI GPUs for 250 epochs with the Adam optimizer~\cite{kingma2014adam}, batch size 6, initial learning rate $10^{-3}$ and weight decay $10^{-8}$. The learning rate was decayed by 0.9 every 10 epochs. Without resampling, each CT slice was cropped by a $256\times256$ window located at the center to focus on the body part. Random cropping of size $16\times128\times128$ was used for data augmentation. Our code is available at \url{https://github.com/HiLab-git/SepNet}.
\subsection{Results of StructSeg 2019}
\par We trained six models with three different SLFs (i.e., SLF1, SLF2 and SLF3) and two loss functions (i.e., $ATH(\alpha=0.5)-L_{Exp}$ and $ATH(\alpha=1)-L_{Exp}$) with our 3D-SepNet. Our algorithms were containerized  with Docker$\footnote{https://www.docker.com/}$ and submitted to organizers of the StructSeg 2019 challenge to get results for the official testing set containing 10 images. For each test case, the algorithm requires at most 120 seconds to generate the output. Quantitative evaluation of the segmentation accuracy was based on Dice Similarity Coefficient ($DSC$) and $95\%$ Hausdorff Distance ($95\%HD$). 
\begin{equation}
    DSC = \frac{2\times TP}{2\times TP+FN+FP}
\end{equation}
where $TP$, $FP$ and $FN$ are true positive, false positive and false negative, respectively. The maximum Hausdorff Distance is the maximum distance of a set to the nearest point in the other set. The maximum Hausdorff Distance from set $X$ to set $Y$ is a maximin function, defined as
 \begin{equation}
     d_H(X,Y)=\max\{ \max\limits_{x\in X}\min_{y\in Y}d(x,y),\max\limits_{y\in Y}\min\limits_{x\in X}d(x,y) \}
 \end{equation}
\par 95\%HD is similar to maximum HD, and it is based on the calculation of the 95th percentile of the distances between boundary points in $X$ and $Y$. The purpose for using this metric is to eliminate the impact of a very small subset of the outliers. Each type of the objective scores of different organs are weighted by their importance weights and then averaged. The 22 annotated OARs with importance weights include: left eye (100), right eye (100), left lens (50), right lens (50), left optical nerve (80), right optical nerve (80), optical chiasma (50), pituitary (80), brain stem (100), left temporal lobes (80), right temporal lobes (80), spinal cord (100), left parotid gland (50), right parotid gland (50), left inner ear (70), right inner ear (70), left middle ear (70), right middle ear (70), left temporomandibular joint (60), right temporomandibular joint (60), left mandible (100), right mandible (100). 
 Eventually, the proposed method achieved weighted average DSC of 80.52\% and 95\%HD of 3.043~mm on the official test set.
 \par For ablation study, as the ground truth of the official testing images in StructSeg 2019 was not publicly available, we did not use any in-house data and randomly split the StructSeg 2019 training set into 40 images for training and the other 10 images for testing, which is referred to as local testing data. 

\subsubsection{Comparison between Intensity Transform and Loss Functions}\label{section:4.1.1}
\begin{figure*}
	\centering
	\includegraphics[width=0.95\linewidth]{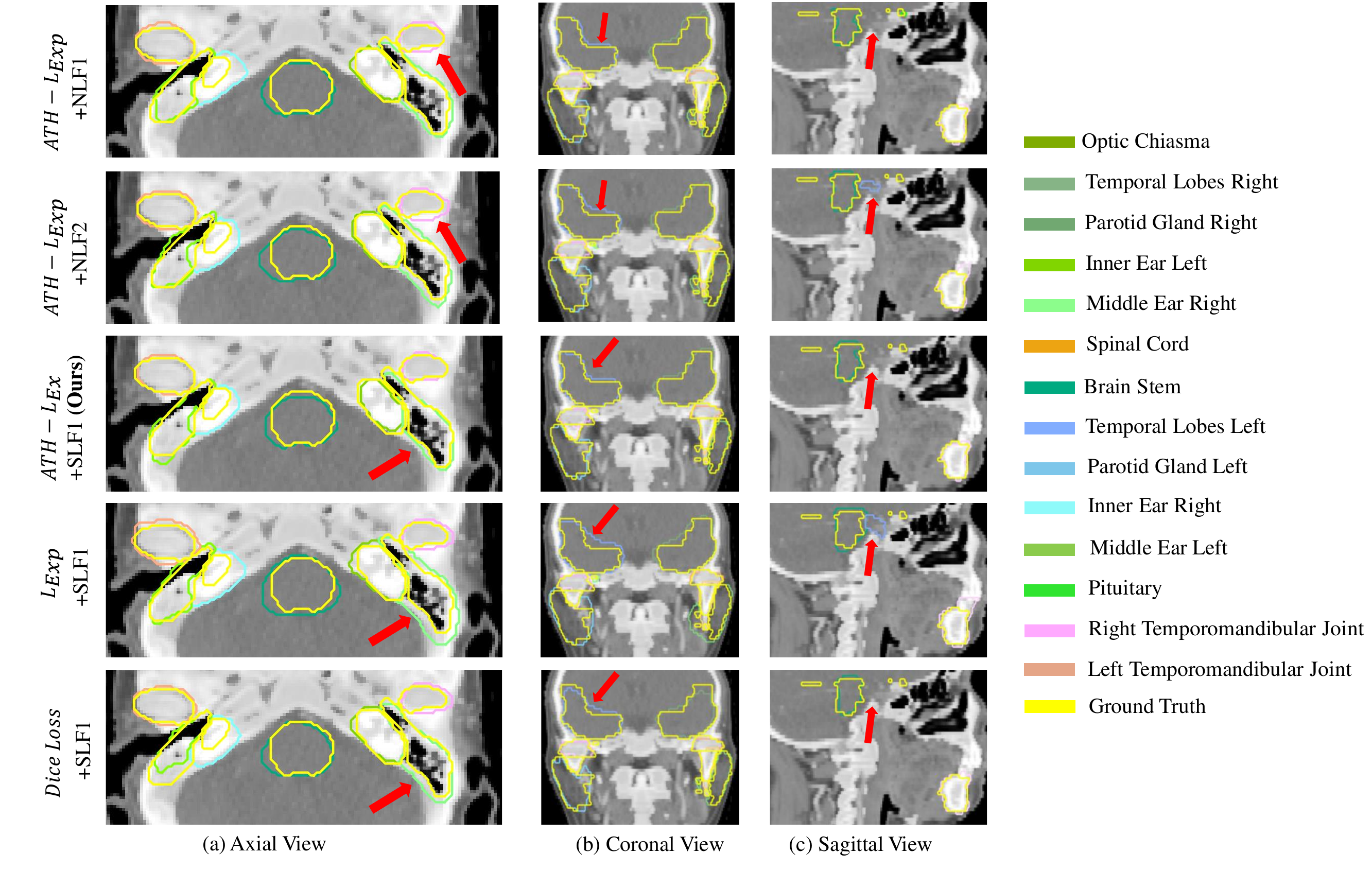}
	\caption{Visual comparison of different intensity transformations and loss functions for segmentation of OARs, where SLF means Segmental Linear Function and NLF represents Naive Linear Function. $L_{Exp}$ is the exponential logarithmic loss and $ATH-L_{Exp}$ represents the combination of our proposed $ATH$ and $L_{Exp}$. Pointed by the red arrows in three views, it could be observed that models trained with SLF could distinguish OARs boundary, i.e. inner ears and temporal lobes more accurately than models trained with NLFs. What's more, compared with $L_{Exp}$ and Dice loss, $ATH-L_{Exp}$ has network pay more attention to hard pixels like boundary in easy or large regions, thus achieving more accurate results. }
	\label{fig:trans_and_loss_compare}
\end{figure*}
\begin{table*}
	\caption{Quantitative comparison between our 3D-SepNet and 3D-UNet based versions. These networks were trained with $ATH(\alpha=0.5)-L_{Exp}$ and SLF1. Small organs contain lens, optical nerves, optical chiasma, pituitary and inner ears. The best result is in bold font.}
	\par
	\centering
	\scalebox{0.75}{
		\centering
		\begin{tabular}{c |c c c c c| c c c c c}
			\hline
			\textbf{Metric}
			&
			\multicolumn{4}{c}{DSC (mean$\pm$std \%)$\uparrow$} &&&
			\multicolumn{4}{c}{95\%HD (mean$\pm$std mm)$\downarrow$}
			\\
			\cline{1-11}
			\textbf{Network}& 
			\textbf{3D-SepNet (ours)} & 3D-UNet & 3D-UNet-Res & 3D-Unet-SE &&&
			\textbf{3D-SepNet (ours)} & 3D-UNet & 3D-Unet-Res & 3D-Unet-SE
			\\ \hline
			Brain Stem & $\bm{89.7{\pm}1.7}$ & 86.9$\pm$1.8 & 87.4$\pm$2.5 & 80.8$\pm$2.7 & & & $\bm{2.79{\pm}0.27}$ & 3.30$\pm$0.28 & 3.5$\pm$1.3 & 4.08$\pm$0.48 \\
			Eye L & 88.6$\pm$3.4 & 88.5$\pm$2.9 & $\bm{88.7{\pm}2.2}$ & 49.9$\pm$18.4 & & & 2.47$\pm$0.31 & $\bm{2.36{\pm}0.25}$ & 2.45$\pm$0.55 & 8.43$\pm$3.13 \\ 
			Eye R & 87.3$\pm$2.5 & 87.4$\pm$4.4 & 88.1$\pm$4.4 & 50.1$\pm$19.9 & & & 2.56$\pm$0.20 & 2.42$\pm$0.38 & 2.62$\pm$0.48 & 8.17$\pm$3.33 \\ 
			Lens L & $\bm{81.5{\pm}6.9}$ & 75.2$\pm$5.2 & 75.6$\pm$4.4 & 7.1$\pm$13.4 & & & 2.37$\pm$0.38 & 2.44$\pm$0.40 & $\bm{1.99{\pm}1.48}$ & 10.09$\pm$5.71 \\ 
			Lens R & $\bm{75.4{\pm}13.0}$ & 71.8$\pm$16.0 & 69.9$\pm$15.6 & 7.3$\pm$19.4 & & & 2.47$\pm$0.45 & 2.57$\pm$0.64 & $\bm{1.85{\pm}0.93}$ & 9.56$\pm$3.66 \\
			Opt Nerve L & $\bm{68.0{\pm}10.7}$ & 66.9$\pm$12.6 & 59.0$\pm$15.4 & 48.2$\pm$9.5 & & & 2.90$\pm$0.78 & $\bm{2.89{\pm}0.74}$ & 3.76$\pm$2.0 & 5.47$\pm$1.12 \\ 
			Opt Nerve R & $\bm{66.3{\pm}11.7}$ & 60.9$\pm$12.6 & 57.9$\pm$18.6 & 46.0$\pm$15.8 & & & $\bm{2.93{\pm}0.77}$ & 3.33$\pm$1.02 & 3.74$\pm$2.02 & 5.68$\pm$1.85 \\ 
			Opt Chiasma & $\bm{56.6{\pm}8.7}$ & 52.2$\pm$10.9 & 40.5$\pm$17.7 & 40.7$\pm$10.5 & & & $\bm{3.83{\pm}0.87}$ & 4.28$\pm$1.10 & 4.69$\pm$1.57 & 3.99$\pm$0.65 \\ 
			Temporal Lobes L & $\bm{86.2{\pm}3.3}$ & 84.7$\pm$3.7 & 85.9$\pm$3.2 & 72.8$\pm$4.7 & & & $\bm{4.31{\pm}0.85}$ & 4.80$\pm$1.07 & 6.62$\pm$1.66 & 8.81$\pm$1.55 \\ 
			Temporal Lobes R & $\bm{86.4{\pm}2.8}$ & 85.4$\pm$4.1 & 85.6$\pm$2.1 & 72.3$\pm$6.3 & & & $\bm{4.38{\pm}0.77}$ & 4.83$\pm$1.28 & 7.06$\pm$1.19 & 8.92$\pm$1.45 \\ 
			Pituitary & $\bm{66.1{\pm}16.3}$ & 65.9$\pm$19.8 & 61.9$\pm$14.4 & 54.2$\pm$14.6 & & & 2.92$\pm$1.01 & 3.01$\pm$1.25 & $\bm{2.9{\pm}0.82}$ & 3.14$\pm$0.77 \\ 
			Parotid Gland L & $\bm{85.7{\pm}3.0}$ & 85.5$\pm$3.8 & 84.9$\pm$5.2 & 57.1$\pm$12.2 & & & $\bm{3.71{\pm}0.64}$ & 3.96$\pm$0.69 & 4.09$\pm$1.73 & 8.75$\pm$1.47 \\ 
			Parotid Gland R & $\bm{87.3{\pm}3.0}$ & 87.0$\pm$3.0 & 87.2$\pm$3.4 & 61.0$\pm$12.9 & & & 3.49$\pm$0.66 & 3.91$\pm$1.44 & $\bm{3.31{\pm}0.9}$ & 8.24$\pm$2.12 \\
			Inner Ear L & 80.3$\pm$6.9 & 78.0$\pm$6.0 & $\bm{80.8{\pm}5.0}$ & 65.2$\pm$11.0 & & & $\bm{2.85{\pm}0.62}$ & 2.90$\pm$0.79 & 3.1$\pm$0.7 & 4.33$\pm$1.17 \\ 
			Inner Ear R & $\bm{80.1{\pm}6.8}$ & 76.2$\pm$7.3 & 79.6$\pm$5.6 & 64.1$\pm$10.8 & & & $\bm{2.77{\pm}0.45}$ & 3.01$\pm$0.67 & 3.06$\pm$1.13 & 4.71$\pm$1.65 \\
			Mid Ear L & 82.6$\pm$5.7 & 81.8$\pm$6.6 & $\bm{83.1{\pm}4.7}$ & 56.4$\pm$12.8 & & & $\bm{3.33{\pm}0.87}$ & 4.40$\pm$2.01 & 4.42$\pm$2.9 & 8.46$\pm$3.4 \\ 
			Mid Ear R & $\bm{78.3{\pm}11.9}$ & 77.1$\pm$10.1 & 78.2$\pm$9.8 & 54.2$\pm$14.9 & & & 3.54$\pm$1.24 & $\bm{3.51{\pm}0.84}$ & 4.01$\pm$1.85 & 7.51$\pm$1.87 \\
			TM Joint L & $\bm{75.7{\pm}9.5}$ & 74.9$\pm$8.3 & 72.0$\pm$8.4 & 54.2$\pm$4.0 & & & $\bm{2.85{\pm}0.70}$ & 3.02$\pm$0.69 & 3.47$\pm$0.92 & 5.87$\pm$0.87 \\ 
			TM Joint R & $\bm{77.2{\pm}7.8}$ & 76.6$\pm$5.4 & 74.2$\pm$5.6 & 56.7$\pm$15.8 & & & $\bm{2.63{\pm}0.51}$ & 2.67$\pm$0.39 & 3.05$\pm$0.13 & 5.21$\pm$1.87 \\ 
			Spinal Cord & 83.0$\pm$4.0 & $\bm{83.7{\pm}3.3}$ & 82.9$\pm$3.5 & 46.5$\pm$21.6 & & & 2.78$\pm$0.33 & 2.73$\pm$0.41 & $\bm{2.06{\pm}0.58}$ & 6.87$\pm$2.73 \\ 
			Mandible L & 91.1$\pm$2.9 & $\bm{91.5{\pm}2.8}$ & 91.5$\pm$2.9 & 45.6$\pm$17.8 & & & 2.81$\pm$0.45 & $\bm{2.54{\pm}0.45}$ & 3.07$\pm$1.13 & 7.25$\pm$1.69 \\ 
			Mandible R & 91.7$\pm$1.5 & 91.7$\pm$2.1 & $\bm{92.0{\pm}1.8}$ & 47.0$\pm$22.3 & & & 2.70$\pm$0.40 & $\bm{2.61{\pm}0.45}$ & 2.75$\pm$0.57 & 7.46$\pm$2.13 \\ \hline
			\textbf{Average} & $\bm{80.2{\pm}2.3}$ & 78.6$\pm$3.0 & 77.6$\pm$4.1 & 51.7$\pm$9.5 & & & $\bm{3.06{\pm}0.24}$ & 3.25$\pm$0.28 & 3.53$\pm$0.52 & 6.86$\pm$1.29 \\ 
			\textbf{Small Organs Average} & $\bm{71.8{\pm}9.4}$ & 68.4$\pm$ 10.0 & 65.7 $\pm$ 12.1 & 41.6$\pm$13.1 & & & $\bm{2.88{\pm}0.67}$ & 3.05$\pm$0.83 & 3.14$\pm$1.3 & 5.5$\pm$2.07
			\\ \hline
	\end{tabular}}
	\label{tab:net_compare}
\end{table*}
\begin{figure*}
	\centering
	\includegraphics[width=0.65\linewidth]{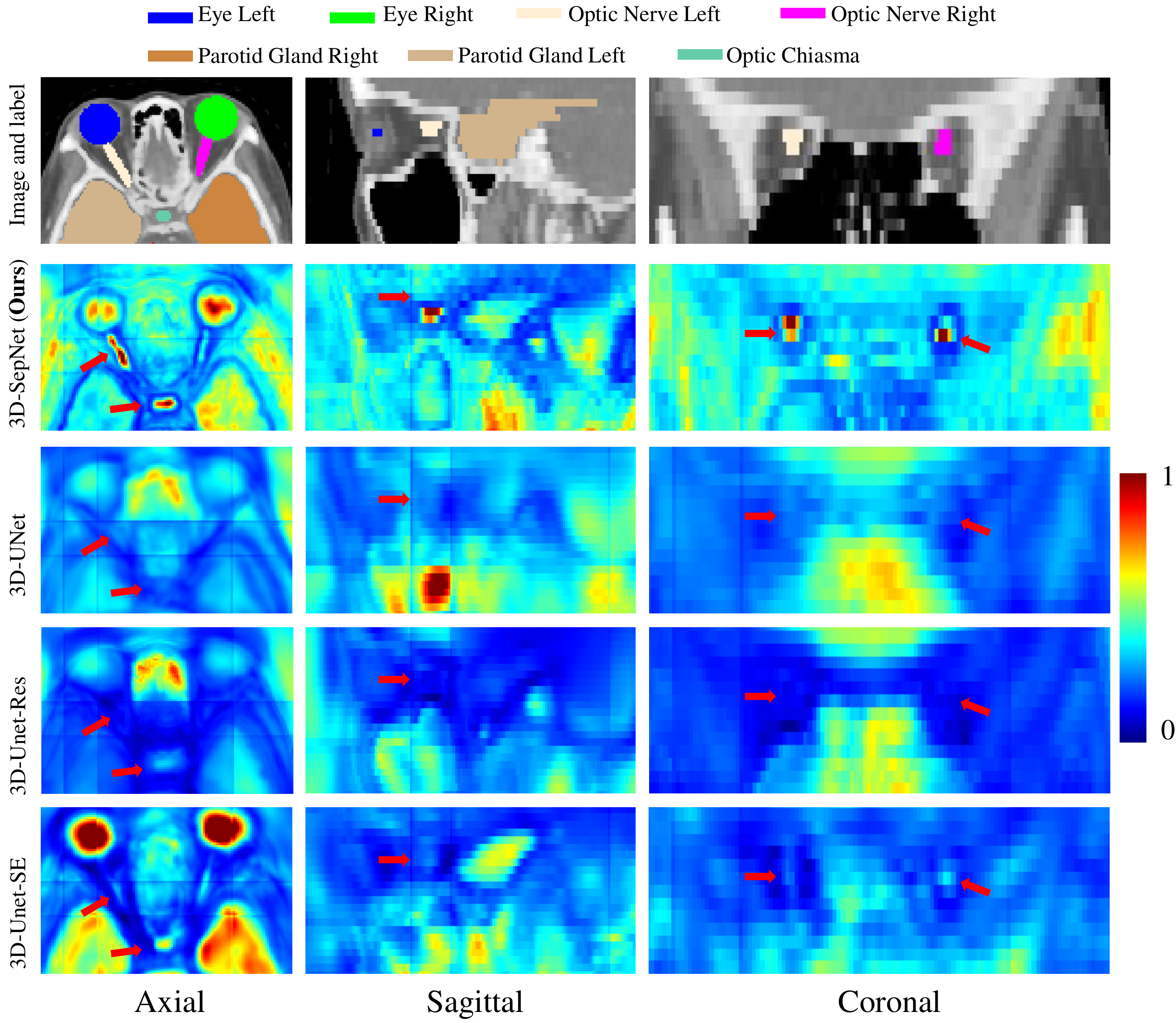}
	\caption{Example of averaged activation maps of different networks. We map the value of each map $m$ from [$m_{min}$, $m_{max}$] to [0, 1] for visualization, where $m_{min}$ and $m_{max}$ are the min and max values of origin map $m$. The red arrows in three views point out that for small organs in large inter-slice, 3D-SepNet is actually benefitted from our separable convolutions and could draw more accurate boundary than the other compared networks.}
	\label{fig:activation_map}
\end{figure*}
\par For quantitative comparison, Table~\ref{tab:trans_and_loss_compare} reports the performance on our local testing set achieved by models trained with 3D-SepNet and mainly two groups: (1) 3 intensity transform functions: SLF1, NLF1 and NLF2 with $ATH(\alpha=0.5)-L_{Exp}$; (2) 3 different loss functions: $ATH(\alpha=0.5)-L_{Exp}$, $L_{Exp}$, Dice loss with SLF1 transform to investigate the effectiveness of our $ATH-L_{Exp}$. 
As shown in Table~\ref{tab:trans_and_loss_compare}, our SLF1 exceeds NLF1/2 at 14 OARs in terms of DSC and 95\%HD.  Due to better visibility of both bones and soft issues obtained by SLFs, model trained with SLFs could learn robust features to determine multiple OARs. In contrast, models taking NLFs as input achieve lower average performance because NLFs cannot differentiate multiple OARs well. Our SLF1 improves Dice by 0.6\% and 0.7\% compared with traditional method NLF1/2 in average, which is a convincing result as this task contains 22 organs.
\par It could be observed that $ATH-L_{Exp}$ outperforms $L_{Exp}$ in most organs like brain stem, optical nerves and inner ear, which verifies that $ATH-L_{Exp}$ will force the network to pay more attention to hard voxels compared with $L_{Exp}$ while still preserve the ability to distinguish the small organs.

\begin{figure*}[!h]
	\centering
	\includegraphics[width=0.9\linewidth]{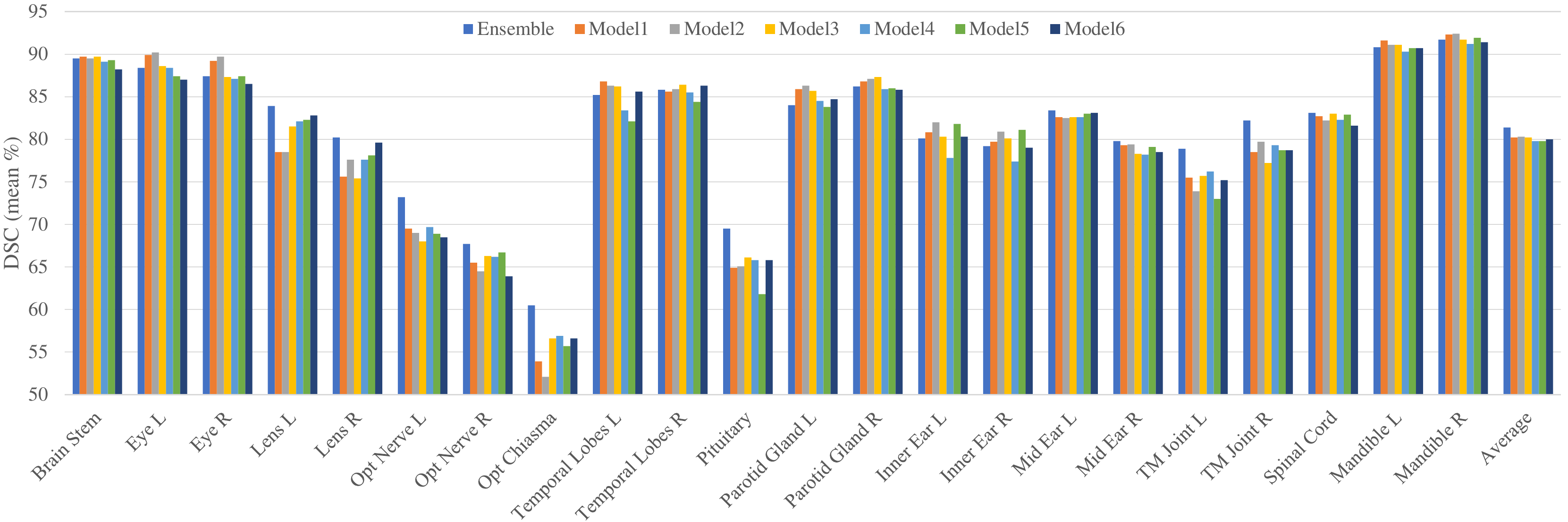}
	\caption{Quantitative comparison of model ensemble with single models for segmentation of OARs. All models are based on 3D-SepNet. Model1/2/3 are trained with $ATH(\alpha=0.5)-L_{Exp}$ and SLF1/2/3 respectively and Model4/5/6 are trained with $ATH(\alpha=1)-L_{Exp}$ and SLF1/2/3 respectively.}
	\label{fig:ensemble_result}
\end{figure*}

\begin{figure*}
	\centering
	\includegraphics[width=.9\linewidth]{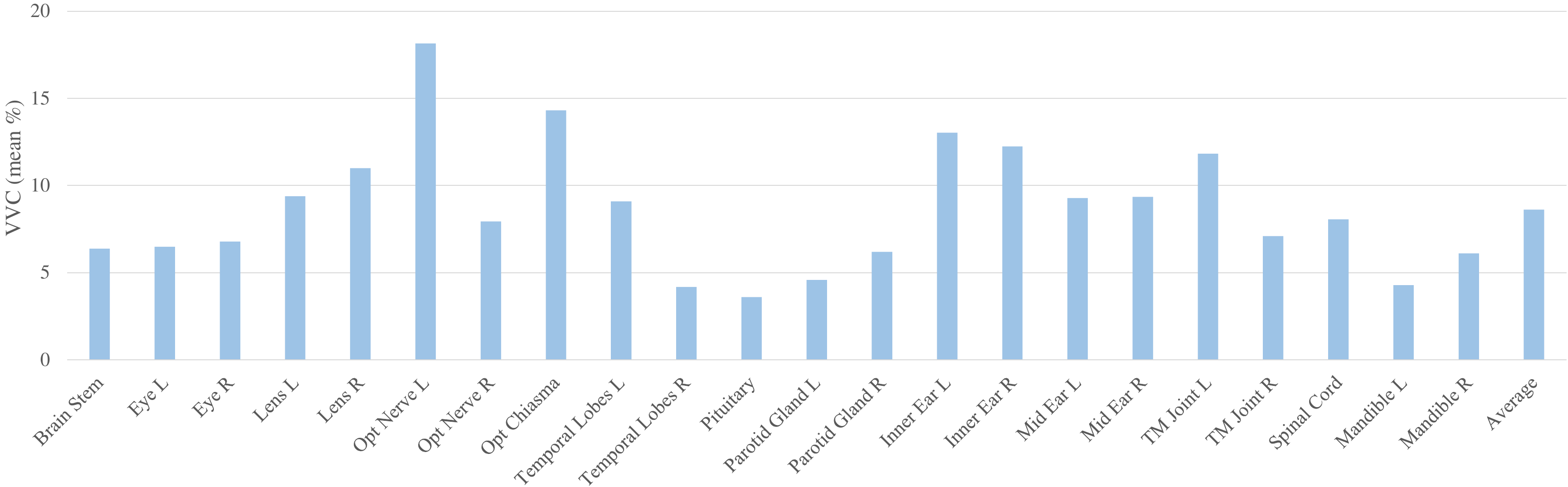}
	\caption{Structure-level uncertainty  in terms of VVC of OARs in the segmentation result. Each VVC is from 6 models used for ensemble and averaged within local test dataset.}
	\label{fig:VVC}
\end{figure*}

\par For qualitative comparison,  Fig.~\ref{fig:trans_and_loss_compare} shows a visual comparison of segmentation results based on above-mentioned five models. It is easy to observe that in all views, the models trained with a wide intensity window of NLF2 failed to recognize the accurate border of brain stem, right temporomandibular joint and left parotid gland due to the low contrast of transformed images. In the contrary, the model trained with SLF1 could distinguish these OARs boundaries effectively, demonstrating the superiority of our proposed intensity transform based on SLFs. 
\par It is also visible that for the hard voxels (near the boundaries of OARs) in easy organs like middle ears and brain stem, segmentations based on $L_{Exp}$ and Dice loss are not very accurate, which is in line with our inference in Section~\ref{section:3.3} that the $L_{Exp}$ may restrain the gradients of hard voxels in easy region. In contrast, $ATH-L_{Exp}$ makes the network focus on these voxels and provides more accurate results.

\subsubsection{Comparison between Networks}

\par We also compare our 3D-SepNet with three variants of 3D-UNet: original one, SE block added and residual connection added. As shown in Table~\ref{tab:net_compare}, with the same loss function and intensity transform function, 3D-UNet variants have a lower performance than 3D-SepNet in small flat organs containing: lens, optical nerves, optical chiasma, pituitary and inner ears that are only present in few slices. The reason could be that after each inter-slice convolution, the boundary along through-plane direction will be blurred. For these flat organs, blurring the through-plane boundary can lead to more reduced segmentation accuracy than those organs that span a larger number of slices where the blurred boundary region occupy a relatively small volume. Thus, our network is more friendly to flat organs than other 3D-UNet versions. To illustrate this point, for each network, we output the activation maps before the last $1\times1$ convolution averaged by channel numbers. We map the value of each map $m$ from [$m_{min}$, $m_{max}$] to [0, 1] for visualization, where $m_{min}$ and $m_{max}$ are the min and max values of origin map $m$. As presented in Fig.~\ref{fig:activation_map}, the averaged activation maps from 3D-SepNet show high response in small organs, e.g.,  fitting well to the optic nerves boundary especially in sagittal and coronal views. In contrast, shape and boundary for the same organs are unclear in the averaged activation maps of other counterparts. What's more, in Table~\ref{tab:net_compare}, it could also be observed that 3D-UNet and 3D-UNet-Res achieve similar performance with 3D-SepNet in organs with large scales in through-plane direction, like spinal cord, mandible, parotid gland and mandible. With about 1/3 of the parameters of 3D-UNet, our 3D-SepNet achieved prominent improvement in both DSC and 95\%HD averagely with the same setting for training, demonstrating the redundancy of usual 3D convolution and that our spatially separable convolution method could deal with anisotropic images more effectively.

\subsubsection{Ensemble and Uncertainty Estimation Results}
Section~\ref{section:4.1.1} shows that it is difficult for a single intensity transform function and a single loss function to achieve the best performance for all OARs, and different SLFs and loss functions are complementary to each other. Therefore, we use an ensemble of them for more robust segmentation.  Our ensemble is based on 6 trained models, containing the combination of three SLFs with loss functions $ATH-L_{Exp}$ of $\alpha=$ 0.5 and 1.0, respectively. Fig.~\ref{fig:ensemble_result} shows a quantitative comparison of six models and their ensemble for segmentation of OARs. It could be observed that the model ensemble improved the average DSC by 1\% for all the OARs and by around 5\% for  small organs such as the lens, the optical chiasma and the pituitary. This demonstrates that our method to take advantage of different complementary intensity transform functions and loss functions effectively improves the segmentation robustness. % with reduced segmentation bias.

\begin{figure*}[!h]
    \centering
    \includegraphics[width=1\linewidth]{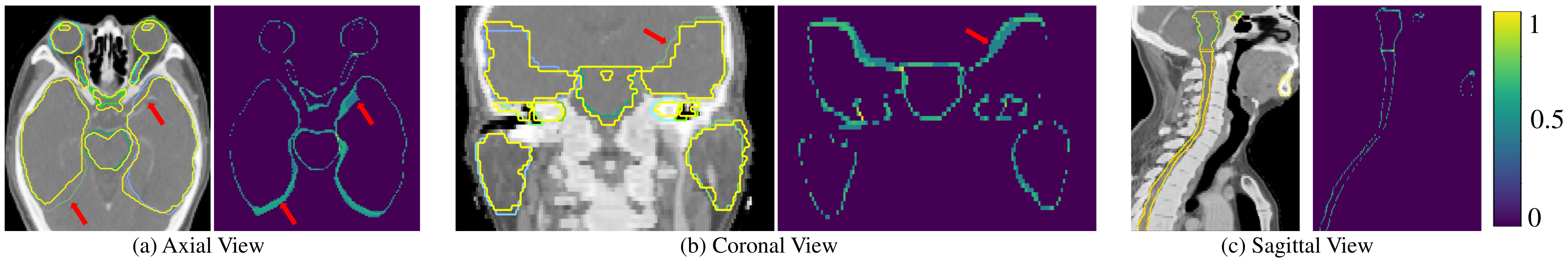}
    \caption{An example of our segmentation uncertainty estimation, encoded by color bar in the left top corner, with the yellow voxels having high uncertainty values and purple voxels having low uncertainty. The left figure in each sub-figure is our ensemble segmentation result. The red arrows point out the major overlap between highly uncertain region and mis-segmentation. }
    \label{fig:uncertain_region}
\end{figure*}
\begin{figure*}[!h]
    \centering
    \includegraphics[width=.95\linewidth]{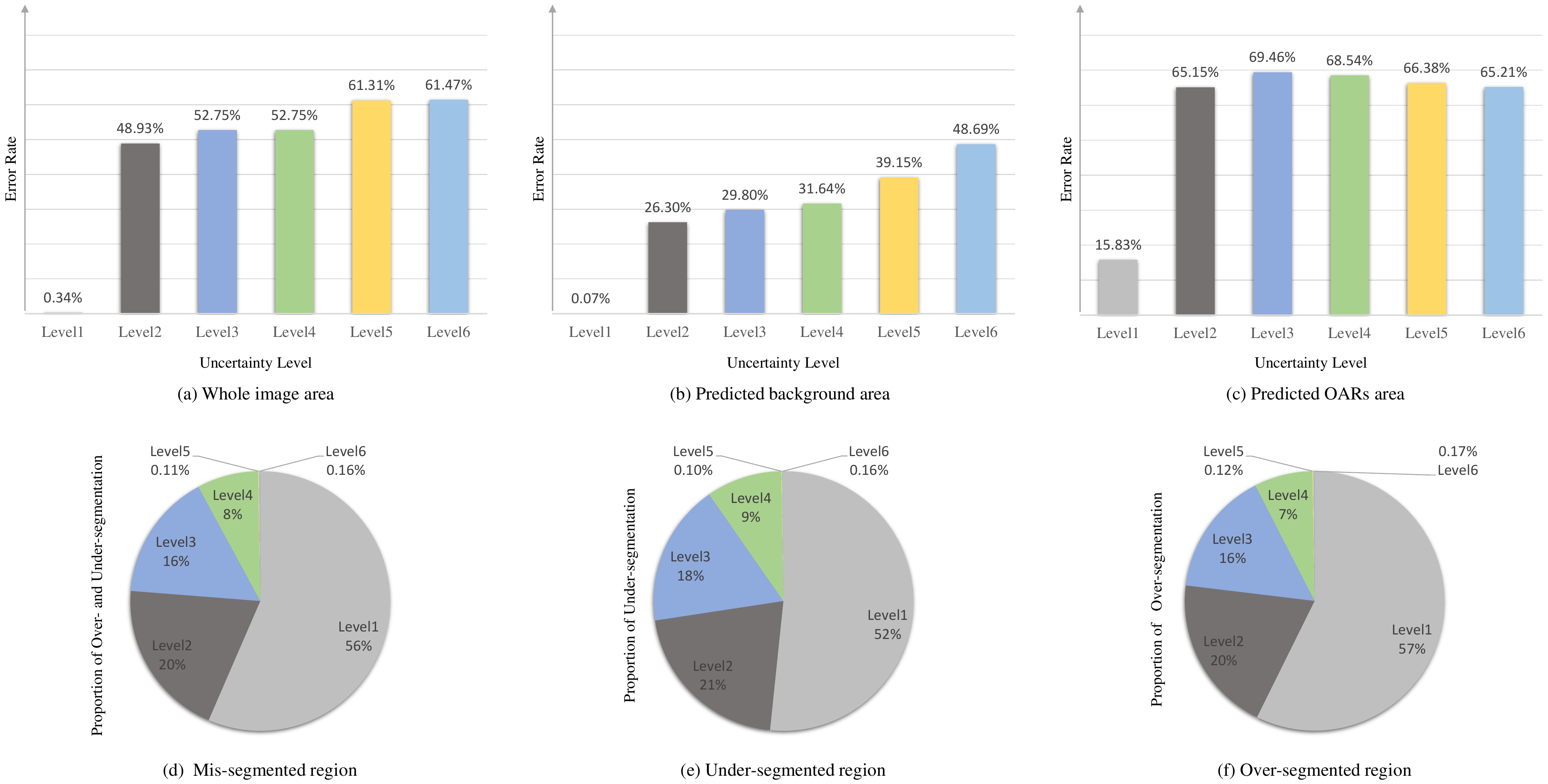}
    \caption{Statistics of segmentation error and voxel-wise uncertainty. (a-c) voxel-wise prediction error rates at different uncertainty levels for the whole image, the predicted background region and the predicted OARs region, respectively. (d-f) Distribution of uncertainty levels in all mis-segmented region, under-segmented region, and over-segmented region, respectively. With 6 levels of uncertainty totally: Level1: 0, Level2: 0.451, Level3: 0.637, Level4: 0.693,  Level5: 0.8675, Level6: 1.011.}
    \label{fig:error_rate}
\end{figure*}
%\subsubsection{Correlation between Uncertainty and Segmentation Error}   
\par As we estimated the voxel-wise uncertainty based on segmentation labels obtained by 6 models using Eq.~\eqref{eq:voxel_uncertain}, we observed 6 levels of voxel-wise uncertainty totally: Level1: 0, Level2: 0.451, Level3: 0.637, Level4: 0.693, Level5: 0.8675, Level6: 1.011. Fig.~\ref{fig:uncertain_region} shows an example of uncertainty information as a result of our ensemble model. In each subfigure, the first image shows the ensemble result compared with the ground truth, and the second image shows uncertainty estimation encoded by a color bar, with the yellow voxels having high uncertainty and purple voxels having low uncertainty. It could be observed that most of the uncertain segmentation belongs to the soft issues and locates near the border of OARs. What's more, the red arrows in Fig.~\ref{fig:uncertain_region} point out the the overlap between highly uncertain regions and mis-segmented areas, showing the uncertainty is highly related to segmentation error. 
\par We also calculated voxel-wise average error rate at each uncertainty level based on different regions: 1) the entire image, 2) predicted background region, and 3) the predicted foreground region of OARs. From the results shown in Fig.~\ref{fig:error_rate} (a, b), it could be observed that for regions with uncertainty Level1 (i.e., all the models obtained the same result), the average error rate is close to 0 for whole image area and the predicted background area. It is mainly due to the large amount of easy-to-recognize voxels in the background. For the predicted OARs area that contains all over-segmented voxels, the average error rate in regions with uncertainty Level1 is much higher and reaches nearly 20\%, as shown in Fig.~\ref{fig:error_rate} (c). With the increase of uncertainty, the results have a steep raise of error rate in all three types of areas. In predicted OARs area, the voxel-wise average error rates for uncertainty Level \{2-6\} are above 65\% while are mainly less than 35\% for uncertainty Level \{2-4\} in predicted background area. It indicates that uncertain region in predicted OARs area is more likely to be mis-segmentation and worth more attention. 
\par What's more, we also calculate the proportion of voxels belonging to each uncertainty level in mis-segmentation for all three types of areas, as shown in Fig.~\ref{fig:error_rate} (d, e, f). It could be observed that the nearly half of voxels in the mis-segmented regions have uncertainty levels between 2 to 6. Therefore, for both predicted background and OARs area, finding and correcting mis-segmented voxels in uncertain region has a large potential for improving the prediction result. 
\begin{figure}
	\centering
	\includegraphics[width=0.8\linewidth]{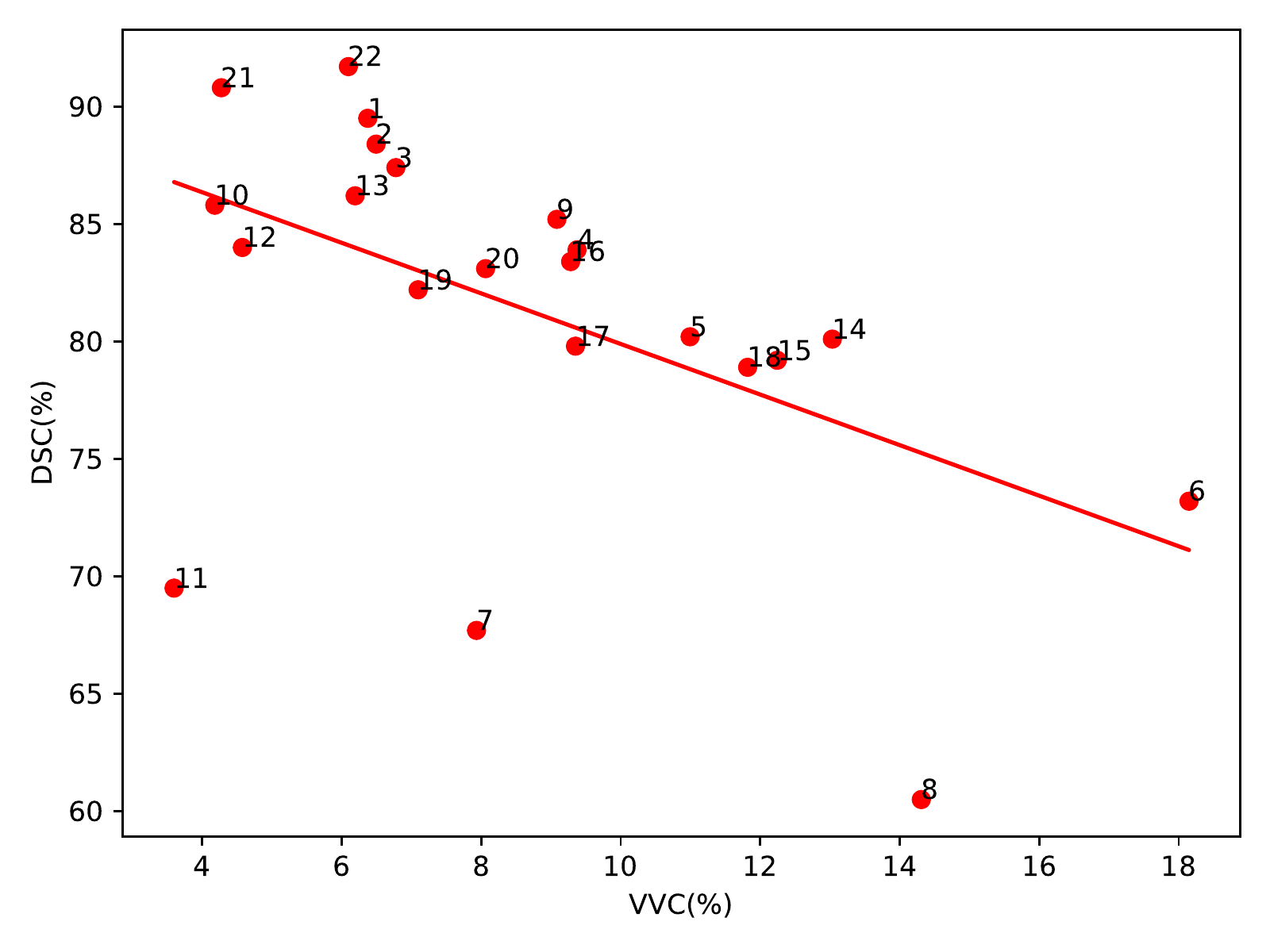}
	\caption{ Joint distribution of DSC and average VVC for each OAR of ensemble result. Each point represents an OAR's DSC and average VVC. With the increase of VVC, the DSC tends to be smaller. The number of each point means the order of OAR: 1-6: Brain Stem, Eye L, Eye R, Lens L, Lens R, Opt Nerve L; 7-12: Opt Nerve R, Opt Chiasma, Temporal Lobes L, Temporal Lobes R, Pituitary, Parotid Gland L; 13-18: Parotid Gland R, Inner Ear L, Inner Ear R, Mid Ear L, Mid Ear R, TM Joint L; 19-22: TM Joint R, Spinal Cord, Mandible L, Mandible R.
	}
	\label{fig:VVC_DSC}
\end{figure}

\begin{table*}[h!]
	\caption{Results of the mixed HAN OARs dataset of different loss functions and transform methods (DSC and 95\%HD (mm)), based on 3D-SepNet.}
	\centering
	\scalebox{0.7}{
		\centering
		\begin{tabular}{c |c c c c c c| c c c c c c}
			\hline
			\textbf{Metric}
			&
			\multicolumn{5}{c}{DSC (mean$\pm$std \%)$\uparrow$} &&&
			\multicolumn{5}{c}{95\%HD (mean$\pm$std mm)$\downarrow$}
			\\
			\cline{1-13}
			\textbf{Transform}& 
			\multicolumn{1}{c|}{NLF1}& NLF2& \multicolumn{4}{|c|}{\textbf{SLF1 (ours)}}& &
			\multicolumn{1}{c|}{NLF1}& NLF2& \multicolumn{3}{|c}{\textbf{SLF1 (ours)}}
			\\
			\cline{1-13}
			\textbf{Loss}&
			\multicolumn{3}{c|}{$\bm{ATH-L_{Exp}}$ \textbf{(ours)}}& 
			\multicolumn{1}{c|}{$L_{Exp}$} & 
			DSC Loss& &
			\multicolumn{4}{c|}{$\bm{ATH-L_{Exp}}$ \textbf{(ours)}}& 
			\multicolumn{1}{c|}{$L_{Exp}$} & 
			DSC Loss
			\\ \hline
			Brain Stem  & 86.6$\pm$3.1 & 87.1$\pm$3.1 & $\bm{87.4{\pm}2.6}$ & 84.3$\pm$3.9 & 86.4$\pm$3.8 & & & 4.04$\pm$1.58 & $\bm{3.73{\pm}1.57}$ & 3.76$\pm$1.47 & 5.48$\pm$1.98 & 4.09$\pm$1.75 \\ 
			Opt Chiasma  & 29.6$\pm$23.6 & 28.3$\pm$24.0 & 29.0$\pm$23.3 & 28.9$\pm$24.8 & $\bm{31.0{\pm}25.7}$ & & & $\bm{9.6{\pm}18.33}$ & 10.38$\pm$17.52 & 10.3$\pm$18.17 & 10.59$\pm$19.26 & 9.92$\pm$18.71 \\ 
			Mandible  & 87.6$\pm$5.8 & $\bm{90.1{\pm}4.5}$ & 90.0$\pm$4.2 & 87.8$\pm$6.5 & 89.1$\pm$7.2 & & & 8.81$\pm$20.4 & $\bm{6.0{\pm}16.09}$ & 6.54$\pm$19.14 & 6.77$\pm$14.83 & 6.58$\pm$15.94 \\ 
			Opt Nerve R  & 62.6$\pm$12.1 & 60.9$\pm$13.1 & 62.4$\pm$13.0 & $\bm{63.3{\pm}13.0}$ & 63.0$\pm$12.1 & & & 4.98$\pm$4.01 & 6.12$\pm$10.22 & 4.43$\pm$3.48 & 4.68$\pm$3.74 & $\bm{4.26{\pm}3.13}$ \\ 
			Opt Nerve L  & 59.8$\pm$12.0 & 61.1$\pm$11.4 & $\bm{62.1{\pm}12.5}$ & 59.9$\pm$13.5 & 58.0$\pm$13.4 & & & 8.42$\pm$14.91 & 6.58$\pm$13.79 & $\bm{4.22{\pm}2.5}$ & 8.05$\pm$22.62 & 7.39$\pm$17.41 \\ 
			Parotid Gland L  & 83.5$\pm$9.5 & 84.1$\pm$7.6 & $\bm{84.7{\pm}5.7}$ & 83.0$\pm$6.4 & 82.3$\pm$10.7 & & & 8.36$\pm$19.22 & 6.77$\pm$11.1 & 6.74$\pm$12.53 & $\bm{5.81{\pm}3.42}$ & 8.51$\pm$17.9 \\ 
			Parotid Gland R  & $\bm{84.7{\pm}4.6}$ & 84.4$\pm$5.5 & 84.6$\pm$4.5 & 83.8$\pm$4.4 & 83.8$\pm$4.8 & & & 5.06$\pm$2.36 & 7.76$\pm$17.64 & $\bm{4.85{\pm}2.39}$ & 5.14$\pm$2.27 & 5.62$\pm$2.86 \\ \hline
			\textbf{Average} & 70.6$\pm$6.7 & 70.8$\pm$6.7 & $\bm{71.5{\pm}6.3}$ & 70.1$\pm$6.7 & 70.5$\pm$7.2 & & & 7.04$\pm$10.28 & 6.76$\pm$12.11 & $\bm{5.83{\pm}7.18}$ & 6.65$\pm$8.48 & 6.62$\pm$10.19 \\ \hline
		\end{tabular}
	}
	\label{tab:trans_and_loss_compare_2}
\end{table*}
\par For evaluation of structure-wise uncertainty, we investigated the relationship between VVC and segmentation accuracy measured by DSC. We calculate VVC for each OAR based on the ensemble of six models and compute the average VVC for each OAR with our local test dataset, as shown in Fig.~\ref{fig:VVC}. From Fig.~\ref{fig:ensemble_result} and Fig.~\ref{fig:VVC}, it can be observed that the left optical nerve has a high structure-level uncertainty with low segmentation accuracy. What's more, we show the joint distribution of average VVC and DSC of ensemble results in Fig.~\ref{fig:VVC_DSC}, where the fitted line shows that DSC tends to be smaller when VVC grows. This demonstrates that a high VVC value can indicate inaccurate segmentation well. The three points far below the fitted line are pituitary, optical nerve right and optical chiasma, %. It represents that for these 3 OARs, different models output predictions which are likely to achieve low accuracy than expectation given the same VCC, 
showing the difficulty of segmenting these OARs. 

\subsection{Results of Mixed HAN OAR Dataset}

\begin{figure}
    \centering
    \includegraphics[width=1\linewidth]{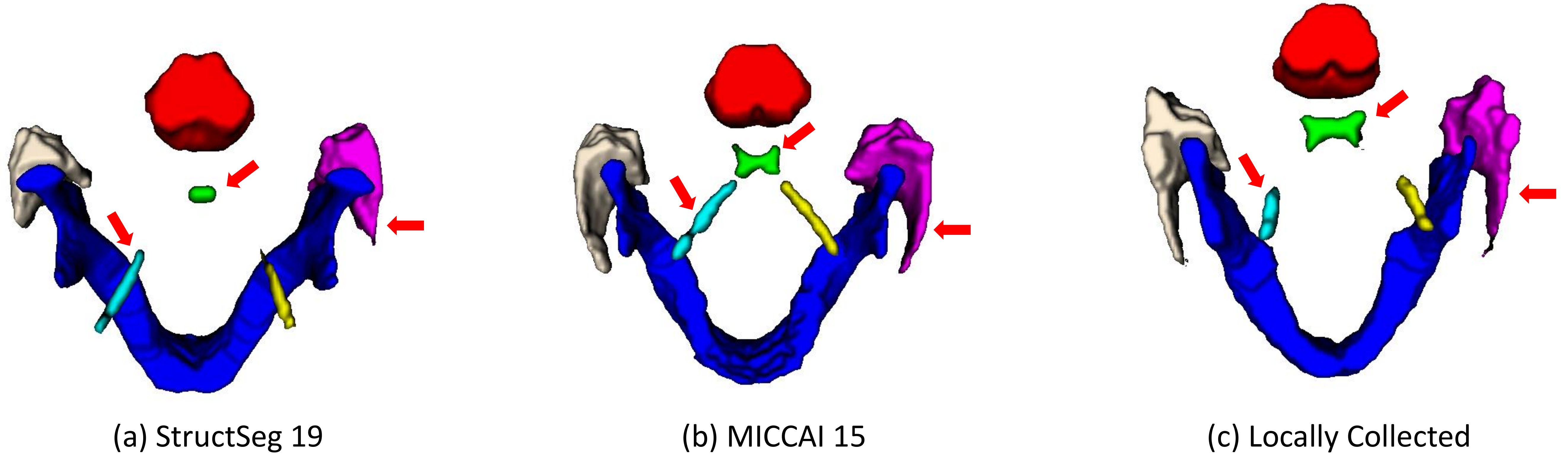}
    \caption{Examples of labeling style from three datasets. The major labeling style differences are pointed out by the red arrows.}
    \label{fig:label_style}
\end{figure}

For further investigation, we applied our methods on a mixed dataset of HAN CT scans of 165 patients: 50 from StructSeg 2019, 48 from MICCAI 2015 Head and Neck challenge and 67 patients collected locally. We segment 7 organs that were annotated in all of them: brain stem, optical chiasma, mandible, optical nerve right/left, parotid gland right/left. However, as shown in Fig.~\ref{fig:label_style}, the labeling style of each dataset is largely different especially in small organs containing optical chiasma and nerves, which makes it a challenging task. The StructSeg 2019 and our locally collected dataset had an inter-slice spacing around 3~mm and intra-slice spacing around 1~mm. The MICCAI 2015 Head and Neck challenge dataset had an intra-slice spacing in the range of 0.76~mm to 1.27~mm, and inter-slice spacing from 1.25~mm to 3~mm. These CT scans  were interpolated to an output voxel size of $3\times1\times1$mm for consistency, and randomly split to 127 for training and 38 for testing. % we also evaluated the results in DSC and 95$\%$HD.
%\subsubsection{Experiment Results}
%\par \textbf{Comparison between Intensity Transform and Loss Functions:} 

Table~\ref{tab:trans_and_loss_compare_2} compares the results of different loss functions and intensity transformation functions based on 3D-SepNet. Due to the different labeling style of each dataset, it is hard to obtain an ideal model for every patient, leading to unstable performance for small OARs like optical chiasma. However, $ATH-L_{Exp}$ with $SLF$ still achieves the best DSC and 95\%HD, proving the robustness of our methods.
In addition, Table~\ref{tab:net_compare_2} shows that compared with other three variants 3D-UNet, our 3D-SepNet also achieves prominent improvements of optical chiasma/nerves especially in terms of DSC. It further illustrates the applicability of our network to handle small organs.
\begin{table*}
\caption{Quantitative comparison between our 3D-SepNet and 3D-UNet based versions on the mixed HAN OARs dataset. These networks were trained with $ATH(\alpha=0.5)-L_{Exp}$ and SLF1.}
\par
\centering
\scalebox{0.7}{
      \centering
    \begin{tabular}{c |c c c c c| c c c c c}
    \hline
        \textbf{Metric}
        &
        \multicolumn{4}{c}{DSC (mean$\pm$std \%)$\uparrow$} &&&
        \multicolumn{4}{c}{95\%HD (mean$\pm$std mm)$\downarrow$}
        \\
        \cline{1-11}
        \textbf{Network}& 
        \textbf{3D-SepNet (ours)} & 3D-UNet & 3D-UNet-Res & 3D-Unet-SE &&&
        \textbf{3D-SepNet (ours)} & 3D-UNet & 3D-Unet-Res & 3D-Unet-SE
        \\ \hline
        Brain Stem  & $\bm{87.4{\pm}2.6}$ & 86.7$\pm$3.4 & 85.7$\pm$3.8 & 86.1$\pm$3.7 & & & $\bm{3.76{\pm}1.47}$ & 3.83$\pm$1.56 & 4.91$\pm$1.86 & 4.03$\pm$1.90 \\ 
        Opt Chiasma  & $\bm{29.0{\pm}23.3}$ & 21.8$\pm$21.9 & 22.9$\pm$22.6 & 24.2$\pm$23.4 & & & $\bm{7.36{\pm}6.62}$ & 8.4$\pm$4.05 & 8.35$\pm$3.55 & 7.78$\pm$3.41 \\ 
        Mandible  & $\bm{90.0{\pm}4.2}$ & 89.5$\pm$6.5 & 89.8$\pm$5.4 & 89.4$\pm$5.1 & & & 6.54$\pm$19.14 & 6.26$\pm$17.06 & $\bm{5.95{\pm}14.75}$ & 6.07$\pm$15.17 \\ 
         Opt Nerve R  & $\bm{62.4{\pm}13.0}$ & 61.8$\pm$13.3 & 61.8$\pm$10.3 & 59.9$\pm$12.4 & & & $\bm{3.82{\pm}3.42}$ & 4.27$\pm$3.12 & 4.04$\pm$2.78 & 4.15$\pm$2.68 \\
        Opt Nerve L  & $\bm{62.1{\pm}12.5}$ & 59.6$\pm$12.4 & 59.9$\pm$11.5 & 58.4$\pm$11.6 & & & $\bm{3.59{\pm}2.5}$ & 4.12$\pm$2.0 & 4.06$\pm$2.16 & 4.54$\pm$2.27 \\ 
        Parotid Gland L  & $\bm{84.7{\pm}5.7}$ & 83.8$\pm$6.9 & 84.2$\pm$6.3 & 83.6$\pm$7.1 & & & $\bm{5.99{\pm}11.52}$ & 7.15$\pm$13.13 & 6.06$\pm$2.81 & 7.34$\pm$3.16 \\ 
        Parotid Gland R  & 84.6$\pm$4.5 & $\bm{84.9{\pm}4.5}$ & 84.4$\pm$4.3 & 84.2$\pm$4.6 & & & $\bm{4.85{\pm}2.39}$ & 7.49$\pm$17.61 & 7.97$\pm$12.47 & 7.71$\pm$16.82 \\ \hline
        \textbf{Average} & $\bm{71.5{\pm}6.3}$ & 69.7$\pm$6.1 & 69.8$\pm$5.8 & 69.4$\pm$6.2 & & & $\bm{5.12{\pm}6.72}$ & 5.93$\pm$7.0 & 5.90$\pm$4.06 & 5.94$\pm$5.08 
        \\ \hline
    \end{tabular}}
\label{tab:net_compare_2}
\end{table*}

\section{Discussion and Conclusion}
To segment multiple OARs with different sizes from CT images with low contrast and anisotropic resolution, we propose a novel framework consisting of Segmental Linear Function (SLF)-based intensity transform and a 3D-SepNet with hard-voxel weighting for training. For intensity transform, the traditional method of Naive Linear Function (NLF) may not be effective enough because it cannot obtain good visibility for multiple OARs including both soft tissues and bones at the same time. In contrast, our SLF makes multiple OARs have a good visibility at the same time. Considering the clinical fact that radiologists would view scans under different window widths/levels to better delineate different organs, we also applied multiple SLFs to better segment different tissues. Generally, our SLF for intensity transform is also applicable to other tasks, such as segmentation of multiple organs from thoracic or abdominal CT images~\cite{gibson2018automatic}. 
\par To deal with the severe imbalance between large and small OARs, we first used the exponential logarithmic loss ($L_{Exp}$) that weights the OARs by the relative size and  class-level segmentation difficulty. However,  we found that $L_{Exp}$ might limit the performance of CNNs on hard voxels in large or easy OARs. To solve this problem, we proposed a weighting function to have the network focus on the hard voxels and combined it with $L_{Exp}$ in our case, named $ATH-L_{Exp}$. As shown in experimental results, the $ATH-L_{Exp}$ outperformed the linear DSC loss and $L_{Exp}$ for most OARs, like brain stem, eye, lens and parotid gland. With a tunable attention parameter $\alpha>0$, we could further control how much the attention is paid to hard regions. Our weighting function could also be easily combined with other segmentation loss functions. % for further improvements. %, especially with those treating voxels in a certain region equally.
\par We validated our framework with extensive experiments in the HAN OAR segmentation task of StructSeg 2019 challenge, and a mixed HAN dataset from three sources. With about 1/3 of the parameters of 3D-UNet, our 3D-SepNet achieved prominent improvement in both DSC and 95\%HD with the same setting for training, demonstrating the redundancy of usual 3D convolution and that our spatially separable convolution method could deal with anisotropic images more effectively.  The experiments found that our 3D-SepNet outperformed 3D-UNet for some flat organs, and it is because 3D-UNet contain more inter-slice convolution operations than our 3D-SepNet and will blur through-plane boundaries. For these flat organs, blurring the through-plane boundaries makes it harder for accurate segmentation. Therefore, using a relatively small number of inter-slice convolution helps to alleviate this problem and obtains better results for organs that are only present in few axial slices. Some other techniques like group-wise~\cite{krizhevsky2012imagenet} and depth separable convolution~\cite{chollet2017xception} could also be combined with our method to build more lightweight and efficient networks. %Our separated convolution is also applicable to other images owning anisotropic voxel-spacing. 
\par Considering that models trained with different loss functions and SLFs obtained the best performance for different OARs, we use an ensemble of these models by a weighted average. %As demonstrated in Fig.~\ref{fig:ensemble_result}, the model ensemble method could reduce the prediction bias and improve the accuracy prominently. 
Our method won the 3rd place\footnote{http://www.structseg-challenge.org/\#/, and our team is UESTC\_501.} in the HAN OAR segmentation task of StructSeg 2019 challenge, and achieved weighted average DSC of 80.52\% and 95\%HD of 3.043 mm based on the official testing images. The leaderboard shows that the difference between our methods and other top-performing methods is not very large in terms of DSC. This is mainly because that the learderboard reports the results averaged across a set of 22 organs that is a mixture of easy and hard organs. We found that UNet could achieve DSC larger than 80\% in near 11 of them, which could result in limited room for improvement on these easy organs. Therefore, significant improvements in hard organs may not lead to a large improvement on the average result. However, it should be noticed that the average DSC of our method is only 0.14\% lower than that of the second best performing method, and is 0.64\% and 0.95\% higher than those of the 4th and 7th top-performing methods, respectively.
\par Further more, to our best knowledge, this is the first work to investigate the uncertainty of HAN OAR segmentation using CNNs. Our results show that there is high correlation between mis-segmentation and high uncertain regions, leading to more informative segmentation outputs that could be used for refinement for more accurate segmentation results and help radiologists during radiotherapy planning. In the future, it is of interest to further improve the segmentation performance on small organs like optical chiasma and optical nerves, and to leverage the uncertainty information to guide user interactions for refinement in challenging cases.

\section*{Acknowledgements}
This work was supported by the National Natural Science Foundations of China [81771921, 61901084] funding.

\bibliographystyle{elsarticle-num-names}
\bibliography{sample}

\end{document}